\documentclass[%
 reprint,
 superscriptaddress,
 longbibliography,
 amsmath,amssymb,
 aps,
 prx,
floatfix,
]{revtex4-2}
\usepackage[english]{babel}
\usepackage{graphicx}
\usepackage{dcolumn}
\usepackage{bm}
\usepackage{comment}
\usepackage{xcolor}
\usepackage{amsmath}
\usepackage{braket}
\usepackage[normalem]{ulem}
\usepackage{float}
\usepackage{MnSymbol}
\usepackage[colorlinks,citecolor=red,urlcolor=blue,bookmarks=false,hypertexnames=true]{hyperref} 
\usepackage{todonotes}

\usepackage{sty_files/coffee4}
\usepackage{subfiles} 

\usepackage[autostyle]{csquotes}
\MakeOuterQuote{"}
\usepackage{todonotes}

\renewcommand{\d}{{\rm d}}

\begin{document}

\preprint{APS/123-QED}

\title{Universal Logical Quantum Photonic Neural Network Processor via Cavity-Assisted Interactions}

\author{Jasvith Raj Basani}
\affiliation{Department of Electrical and Computer Engineering, Institute for Research in Electronics and Applied Physics, and Joint Quantum Institute, University of Maryland, College Park, MD 20742, USA}
\author{Murphy Yuezhen Niu}
\affiliation{Department of Computer Science, University of California, Santa Barbara, CA 93106, USA}
\affiliation{Google Quantum AI, Venice, California 90291, USA}
\affiliation{Department of Computer Science, Joint Center for Quantum Information and Computer Science, University of Maryland, College Park, MD 20742, USA}

\author{Edo Waks}
\affiliation{Department of Electrical and Computer Engineering, Institute for Research in Electronics and Applied Physics, and Joint Quantum Institute, University of Maryland, College Park, MD 20742, USA}

\date{\today}

\begin{abstract}
Encoding quantum information within bosonic modes offers a promising direction for hardware-efficient and fault-tolerant quantum information processing. However, achieving high-fidelity universal control over the bosonic  degree of freedom using native photonic hardware remains a challenge. Here, we propose an architecture to prepare and perform logical quantum operations on arbitrary multimode multi-photon states using a quantum photonic neural network. Central to our approach is the optical nonlinearity, which is realized through strong light-matter interaction with a three-level $\Lambda$ atomic system. The dynamics of this interaction are confined to the single-mode subspace, enabling the construction of high-fidelity quantum gates. This nonlinearity functions as a photon-number selective phase gate, which facilitates the construction of a universal gate set and serves as the element-wise activation function in our neural network architecture. Through numerical simulations, we demonstrate the versatility of our approach by executing tasks that are key to logical quantum information processing. The network is able to deterministically prepare a wide array of multimode multi-photon states, including essential resource states. We also show that the architecture is capable of encoding and performing logical operations on bosonic error-correcting codes. Additionally, by adapting components of our architecture, error-correcting circuits can be built to protect bosonic codes. The proposed architecture paves the way for near-term quantum photonic processors that enable error-corrected quantum computation, and can be achieved using present-day integrated photonic hardware. 
\end{abstract}

\maketitle

\section{Introduction}

Encoding quantum information within bosonic modes has found a number of applications in quantum computation, metrology, and communication. In particular, multimode multi-photon states have been extremely useful in the development of bosonic error-correcting codes to aid fault-tolerant quantum computation~\cite{chuang1997bosonic, niu2018hardware, niu2018qudit, grimsmo2020quantum, albert2018performance, michael2016new}. These error-correcting codes can be tailored to be hardware-efficient~\cite{niu2018hardware, niu2018qudit, krastanov2021room}, and have recently been shown to reach the break-even point for error-corrected quantum computation~\cite{ofek2016extending, hu2019quantum, ni2023beating}. Multimode bosonic states have also been shown to achieve improved resolution and parameter sensitivity in the case of metrology and sensing~\cite{oszmaniec2016random, leibfried2003quantum, nagata2007beating}. Quantum communication protocols utilizing bosonic states are also found to have improved fidelity of quantum state transfer~\cite{michael2016new, lami2020bosonic} and entanglement distribution between nodes~\cite{cirac1997quantum}.

A critical capability for harnessing the full potential of multimode multi-photon states is the high fidelity state preparation and universal quantum control. Standard circuit constructions using single and two-qubit gates are not well suited to manipulate arbitrary multimode multi-photon bosonic states because their code-space spans only a fraction of the entire Hilbert space. Moreover, correcting errors such as photon loss requires complex circuitry that uses additional ancillary modes and potentially faulty gates~\cite{divincenzo1996fault, laflamme1996perfect}.  Alternatively, architectures that naturally operate on the photon-number basis have been proposed using weak second and third-order optical nonlinearities~\cite{krastanov2021room, steinbrecher2019quantum, ewaniuk2023imperfect, niu2018hardware, niu2018qudit}. These nonlinearities generate significant temporal mode distortions, making them extremely challenging to use for controllable manipulation of multi-photon bosonic states~\cite{shapiro2006single, leung2009spectral, gea2010impossibility, he2011transverse, dove2014phase}. Reliably utilizing these optical nonlinearities requires fast time-dependent control of cavity output coupling, which is challenging to implement in practice~\cite{heuck2020photon, heuck2020controlled, choi2017self, yanagimoto2022temporal}. Furthermore, the groups of states and operations that can be implemented by these architectures are limited because they are bound by the symmetry of their respective Hamiltonians. Photon-photon interactions mediated by simple two-level quantum emitters also suffer from time-bandwidth limitations, which impose an upper bound on the fidelity for many quantum operations~\cite{singh2023enhanced, nysteen2017limitations, ralph2015photon, rosenblum2011photon, javadi2015single, jeannic2022dynamical}. We therefore still lack general methods to generate and manipulate multi-photon multi-mode states at scale with high fidelity.

In this paper, we demonstrate a framework to prepare arbitrary multimode multi-photon states, and implement universal and encoded quantum operations using a strong programmable optical nonlinearity. We utilize quantum photonic neural networks, parameterized by phase-shifts to engineer a wide array of logical operations to manipulate quantum information encoded in the complex probability amplitudes of the multimode multi-photon states. The central element of our architecture is the optical nonlinearity, which is implemented by strong light-matter interaction with a three-level $\Lambda$ atomic system~\cite{rosenblum2011photon, rosenblum2016extraction, rosenblum2017analysis}. This nonlinearity is used as an elementwise activation function, and acts as a photon-number selective phase gate. The dynamics of this light-matter interaction is confined to the single-mode subspace, and does not introduce temporal mode distortions which is a common limitation of other optical nonlinearities. 

We demonstrate the versatility of this approach by benchmarking the architecture on tasks representative of logical quantum computation - state and channel preparation. First, we consider the task of resource state preparation, and show that the network is able to prepare a sample of Haar-random multi-photon multimode state to high fidelity. Using the specific example of a 4-photon $N00N$ state, we analyze the impact coherent errors caused by component imperfections on the fidelity of state preparation. Next, we consider the task of preparing unitary quantum channels, which is essential for implementing quantum logic. We demonstrate this by encoding and preparing a universal gate set for bosonic error correcting codes. We use the example of the 5-photon two-mode $\chi^{(2)}$ binomial code, because it \textit{cannot} be prepared using just linear optical elements and a $\chi^{(2)}$ nonlinearity~\cite{niu2018hardware}. These operations can be learned by the network in the encoded basis to a fidelity exceeding $99.9\%$. We also show that the primitive components of our architecture can be adapted to perform non-demolition measurements of the total photon number, and consequently construct error-correcting circuits. Finally, we evaluate the hardware requirements that must be addressed in order to construct the neural network processor, and show that the proposed architecture can be realized using present-day integrated photonic hardware. 

In the next section, we introduce the architecture of the quantum photonic neural network. We describe the nonlinear dynamics implemented by the three-level $\Lambda$ atomic system, and how it can be used to construct deterministic, high-fidelity gates. In sec.~\ref{sec:3}, we discuss the results of our numerical simulations to prepare quantum states and unitary channels. We show that by tiling components of our architecture, it becomes possible to build error correcting circuits that operate on bosonic codes. Lastly, in sec.~\ref{sec:4} we discuss the hardware parameters required to experimentally realize the proposed device on an integrated platform. We conclude with a summary of the main advantages of our proposed architecture, and the future scope of this work.

\begin{figure*}[t!]
    \centering
    \includegraphics[width = \textwidth]{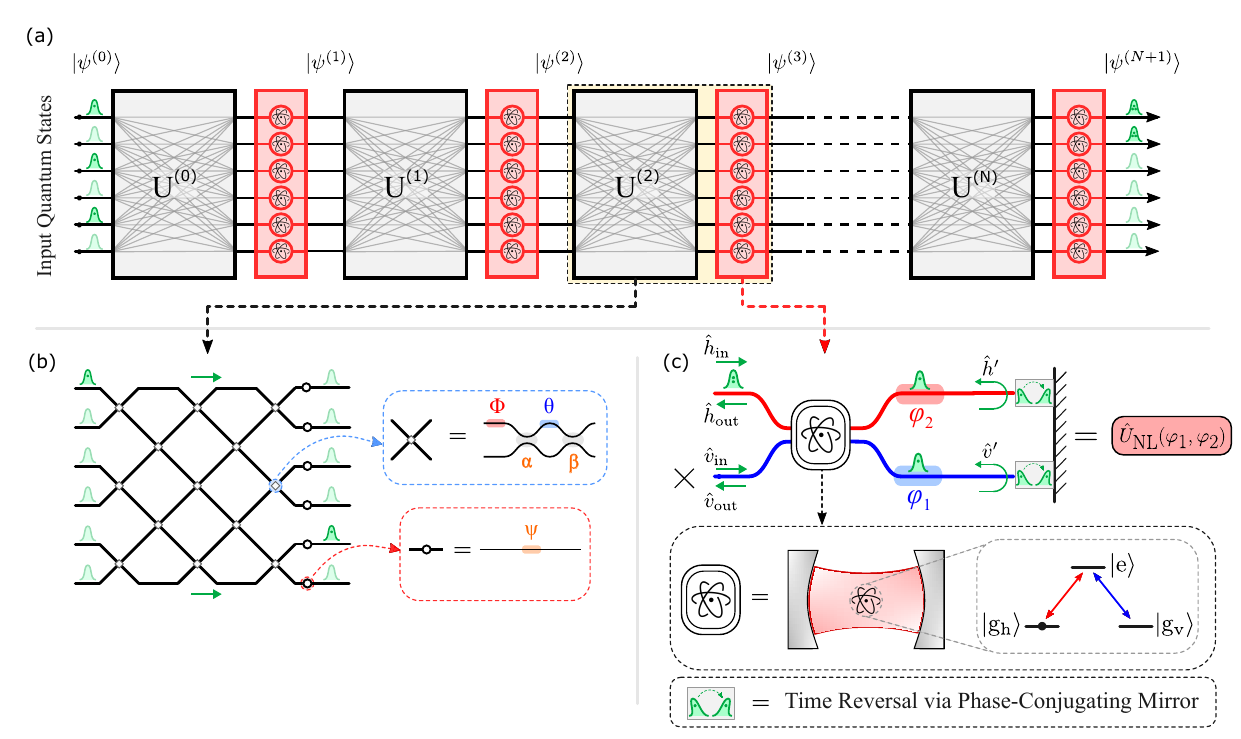}
    \caption{{\fontfamily{cmss}\selectfont\textbf{Schematic of the components of a quantum photonic neural network.} \textbf{(a)} Illustrative representation of a neural network represented as a sequence of $N$ layers. Inputs to this network are multi-photon fock states. Each grey block ($\mathrm{U}^{(i)}$) performs a linear-optical transformation, and the red blocks perform the element-wise nonlinear activation function. \textbf{(b)} Hardware implementation of the linear layer - a multiport interferometer mesh in the \texttt{CLEMENTS} configuration. The insets show the constituent components of the mesh including Mach-Zehnder Interferometers and phase-shifters. \textbf{(c)} Illustrative representation of the schematic to implement the nonlinear activation function. The atom is a three-level $\Lambda$ atomic system coupled to an optical cavity as shown in the inset. Transitions of the three-level atom are coupled to a single optical path, i.e., the $\ket{\mathrm{g}_{\mathrm{h}}} \leftrightarrow \ket{e}$ transition is coupled to the red path and the $\ket{\mathrm{g}_{\mathrm{v}}} \leftrightarrow \ket{e}$ transition is coupled to the blue path. The input and output of the nonlinear element are along the modes $\hat{h}_{\mathrm{in}}$ and $\hat{h}_{\mathrm{out}}$ respectively.}}
    \label{fig:schematic_main}
\end{figure*}

\section{The Neural Network Architecture \label{sec:2}}

 Quantum photonic neural networks have emerged as a promising platform for optical quantum computation. Theoretical proposals have shown that these architectures are excellent at gate synthesis, black-box quantum simulation, and acting as one-way quantum repeaters~\cite{steinbrecher2019quantum, krastanov2021room, ewaniuk2023imperfect}. These approaches, however, utilize weak optical nonlinearities and assume idealized models of the hardware, without accounting for the multimode dynamics of photon-photon interactions. Single mode operation of the nonlinearity places stringent constraints on the hardware, requiring high-speed active switching and control of cavity quality factors. These constraints makes these designs challenging to realize experimentally. In this section, we introduce a hardware model of the quantum photonic neural network based on passive light-matter interactions. This model uses components that have been experimentally realized, combined with with existing photonic hardware.

 A schematic of the neural network architecture that we propose is illustrated in fig.~\ref{fig:schematic_main}(a). It consists of cascaded layers of linear and nonlinear transformations, which are indicated by the grey and red blocks respectively. The network has $M$ distinct spatial modes, to which the inputs are $N$-photon multimode quantum states. These inputs are indistinguishable photons with identical temporal wave-packets. Therefore, the amplitudes of these states are encoded as a complex-valued vector with unit magnitude. The linear transformation layers that apply a unitary operation, denoted by $U^{(i)}$, are implemented by an $M$-port interferometer~\cite{reck1994experimental, clements2016optimal, basani2023self,lee2007balanced,hamerly2024towards}. The nonlinear optical operation, which acts as an element-wise nonlinear activation, is implemented by strong light-matter interaction on each output port of the interferometer.

 The fundamental building block of a programmable multiport interferometer that implements linear optical operations is the 2-port Mach-Zehnder Interferometer (MZI). A schematic of this mesh is shown in fig.~\ref{fig:schematic_main}(b). A single Mach-Zehnder Interferometer (shown in the upper inset) is composed of two beam-splitters and two tunable phase-shifters, parameterized by $(\theta, \phi)$. The interferometer can be programmed to  realize any $2 \times 2$ unitary operation. The beam-splitters in the interferometer are assumed to have a perfect 50:50 splitting ratio. Deviations from ideal splitting ratios in each beam-splitter is parameterized by the angles ($\alpha, \beta$). The transfer function among modes $i, j$ denoted by $T_{i,j}(\theta, \phi)$ of a single Mach-Zehnder Interferometer is given by:
\begin{equation}
    T_{i,j}(\theta,\phi) = i e^{i\theta/2} \begin{bmatrix} e^{i\phi} \sin(\theta/2) & \cos(\theta/2) \\ e^{i\phi} \cos(\theta/2) & -\sin(\theta/2) \end{bmatrix} \label{eq:mzi_t}
\end{equation}
Larger programmable unitary operations $U$ can be implemented by decomposing them into a product of $2 \times 2$ unitary matrices $T_{i,j} (\theta, \phi)$ and phase shifts on the output modes (shown in the lower inset), corresponding to a diagonal matrix $D$. Therefore, an $M \times M$ unitary matrix can be decomposed as $U = D\prod T_{i, j}(\theta, \phi)$. 

The schematic of the hardware that implements the nonlinear optical transformation is depicted in fig.~\ref{fig:schematic_main}(c). This nonlinear gate has two input channels and output channels that serve as propagation channels for orthogonal modes. The output from each port of the linear interferometer serves as the input to the mode denoted by $\hat{h}_{\mathrm{in}}$. The input to $\hat{v}_{\mathrm{in}}$, which encodes the mode orthogonal to $\hat{h}_{\mathrm{in}}$ is always the vacuum state. These modes are used to excite a single three-level $\Lambda$ atomic system, with two ground states $\ket{\mathrm{g}_{\mathrm{h}}}, \ket{\mathrm{g}_{\mathrm{v}}}$ and one excited state $\ket{e}$. Each transition of the atoms couples to only \textit{one} of the modes with equal cooperativity, i.e., pulses in mode $\hat{h}_{\mathrm{in}}$ can excite only the $\ket{\mathrm{g}_{\mathrm{h}}} \leftrightarrow \ket{e}$ transition and pulses in mode $\hat{v}_{\mathrm{in}}$ can excite only the $\ket{\mathrm{g}_{\mathrm{v}}} \leftrightarrow \ket{e}$ transition. The output of the atomic system propagates along modes $\hat{v}'$ and $\hat{h}'$, which each experience independent phase shifts $\varphi_{1}$ and $\varphi_{2}$ respectively. Pulses propagating along these modes are time-reversed via a phase-conjugating mirror and are reflected to interact with the atomic system a second time. The final state from the time-reversed second interaction step then exits the nonlinear gate via the mode $\hat{h}_{\mathrm{out}}$. This model of the activation function therefore acts as a programmable nonlinear phase gate which we denote as $\hat{U}_{\mathrm{NL}}(\varphi_{1}, \varphi_{2})$.

The first step to implement this nonlinear phase gate is the process of photon subtraction. In this configuration, a three-level $\Lambda$ atomic system has been shown to deterministically extract a single photon from an incoming pulse, and move it to an orthogonal mode while flipping its state~\cite{gea2013photon, rosenblum2011photon, rosenblum2016extraction, rosenblum2017analysis, pinotsi2008single, koshino2010deterministic}. The Hamiltonian associated with the interaction of light interacting with a single three-level atomic system trapped in an optical cavity is given by~\cite{gea2013space}:

\begin{multline}
    \hat{H} = -i\hbar g\sqrt{\frac{\kappa}{\pi}} \int \frac{1}{\kappa - i\omega} \Big( \ket{e}\bra{g_{\mathrm{h}}} \hat{a}_{\omega}  \\  
    + \ket{e} \bra{g_{\mathrm{v}}} \hat{b}_{\omega} \Big)  e^{-i(\omega + \delta)t} \d\omega + \mathrm{H.C.}
\end{multline}
where $\hat{a}_{\omega}$ and $\hat{b}_{\omega}$ are the bosonic annihilation operators for the modes $\hat{h}$ and $\hat{v}$ respectively. The cavity is assumed to be perfectly resonant with the atomic transitions, i.e., the detuning $\delta = 0$. The parameter $g$ denotes the coupling strength between the cavity mode and the atomic transition, and $\kappa$ denotes the cavity decay rate. In general, the input to the nonlinear phase gate is an $N$-photon Fock state with temporal mode profiles $\xi(t)$ in the $\hat{h}_{\mathrm{in}}$ mode. In the limit where incident photons are spectrally narrower than the line-width of the cavity ($\kappa \gg \sigma$) and the cavity-enhanced decay rate ($2g^{2}/\kappa \gg \sigma$), the two-mode subtracted state is:
\begin{multline}
    \ket{\psi_{\mathrm{v}} (t, t', t'')} = -\sqrt{\frac{N}{(N - 1)!}} \left( \int_{-\infty}^{t} \d t' \xi(t') \hat{b}^{\dagger}_{t'}   \right) \times \\ 
    \left( \int_{t'}^{\infty} \d t'' \xi(t'') \hat{a}^{\dagger}_{t''} \right)^{N - 1} \ket{0}
    \label{eq:subtracted_state}
\end{multline}
In the limit that $t \to \infty$, we recover the subtracted state derived in ref.~\cite{gea2013photon} (see sec.~\ref{sec:appendix_1} of the Supplementary Information). Eqn.~\eqref{eq:subtracted_state} indicates that the subtracted state is temporally entangled, but occupy orthogonal modes -- a single excitation in mode $\hat{v}'$ and $(N - 1)$ excitations in mode $\hat{h}'$. In this limit, each mode retains the original temporal profile $\xi(t)$ without distortion, meaning that it is confined to the single-mode subspace.

Following the process of photon subtraction, each mode of the two-mode state in eqn.~\eqref{eq:subtracted_state} gains independent phases $(\varphi_{1}, \varphi_{2})$, and is time-reversed. Time-reversal of the two-mode state is performed using an optical phase-conjugating mirror. Mathematically, the phase-conjugating mirror substitutes the forward evolving time scales $(t, t', t'')$ with $(-t, -t', -t'')$ in eqn.~\eqref{eq:subtracted_state}. Perfect time-reversal of an arbitrary pulse using linear optics and time-dependent refractive-index modulation has been proposed and demonstrated in refs.~\cite{yanik2004time, sivan2011time, minkov2018localization, chumak2010all}. Alternatively, nonlinear optical processes such as four-wave mixing has also been demonstrated to perform time-reversal of an optical pulse~\cite{miller1980time, he2002optical, caro1982phase, yariv1979compensation, fisher1983optical, liu2013phase, watanabe1993compensation, pepper1980compensation}. This time-reversed state is reflected and interacts with the three-level $\Lambda$ system a second time. 

Finally, the two-mode time-reversed state interacts with the $\Lambda$ atomic system again, and is added into a single-mode with a global nonlinear phase. This interaction step emulates the process of photon addition, where the time-reversed state interacts with the atom in the $\ket{\mathrm{g}_{\mathrm{v}}}$ state. The mode $\hat{h}'$ cannot excite the $\ket{\mathrm{g}_{\mathrm{v}}} \leftrightarrow \ket{e}$ transition, and therefore propagates unperturbed. The single photon in mode $\hat{v}'$, on the other hand, excites the atom, and is deterministically added to the the pulse in $\hat{h}'$ which propagates into the output mode $\hat{h}_{\mathrm{out}}$ (see sec.~\ref{sec:appendix_1} of the Supplementary Information). This outputs the original $N$-photon fock state with a global nonlinear phase. Therefore, the nonlinear phase gate described in fig.~\ref{fig:schematic_main}(c) performs the following transformations on an input $N$-photon fock state:
\begin{equation}
    \ket{N} \mapsto e^{i\varphi_{1}}e^{i(N - 1)\varphi_{2}} \ket{N}
    \label{eq:nl_transformation}
\end{equation}

This photon-number selectivity enables the construction of multi-qubit entangling gates. In the dual rail basis, along with single qubit gates, two-qubit controlled phase gates are required to realize a universal gate set. The function implemented by the nonlinear phase gate $\hat{U}_{\mathrm{NL}}(\varphi_{1}, \varphi_{2})$ transforms the superposition state $c_{0}\ket{0} + c_{1}\ket{1} + c_{2}\ket{2}$ into the state  $c_{0}\ket{0} + c_{1}e^{i\varphi_{1}}\ket{1} + c_{2}e^{i(\varphi_{1} + \varphi_{2})}\ket{2}$. When $\varphi_{1} = 0, \varphi_{2} = \pi$, the output state is $c_{0}\ket{0} + c_{1}\ket{1} - c_{2}\ket{2}$, which is required to implement a controlled-phase gate~\cite{knill2001scheme, ralph2015photon}. The fidelity of this nonlinear phase gate is discussed in sec.~\ref{sec:appendix_2} of the Supplementary Information. This optical nonlinearity can be extended to construct general photon-number selective arbitrary phase gates for $N > 2$ incident photons. The fidelity of this phase gate and the photon-number selective. This model of the nonlinearity is discussed in sec.~\ref{sec:appendix_3} of the Supplementary Information.

\section{Results \label{sec:3}}

We benchmark the performance of the proposed architecture by numerically simulating tasks such as state preparation, encoding logical information, and realizing logical gates. We also analyze the performance of the network under the influence of realistic hardware errors. The linear layers of our architecture are implemented by a programmable multiport interferometer in the \texttt{CLEMENTS} configuration, parameterized by $(\Vec{\theta}, \Vec{\phi})$. The nonlinear activation is implemented by the programmable nonlinear phase gate described in sec.~\ref{sec:2}, and is parameterized by $(\Vec{\varphi}_{1}, \Vec{\varphi}_{2})$. The phases for both the linear and nonlinear layers are initialized uniformly at random. The digital differentiable model was programmed using open-source automatic differentiation tools with the \texttt{JAX} library, and numerically simulated to optimize the phases of the network. The network was trained using the \texttt{ADAM} optimizer~\cite{kingma2014adam}, with the learning rate annealed from 0.025 to 0.001 on the \texttt{NVIDIA A100} GPU.

\subsection{State Preparation}

\begin{figure}
    \centering
    \includegraphics[width = \columnwidth]{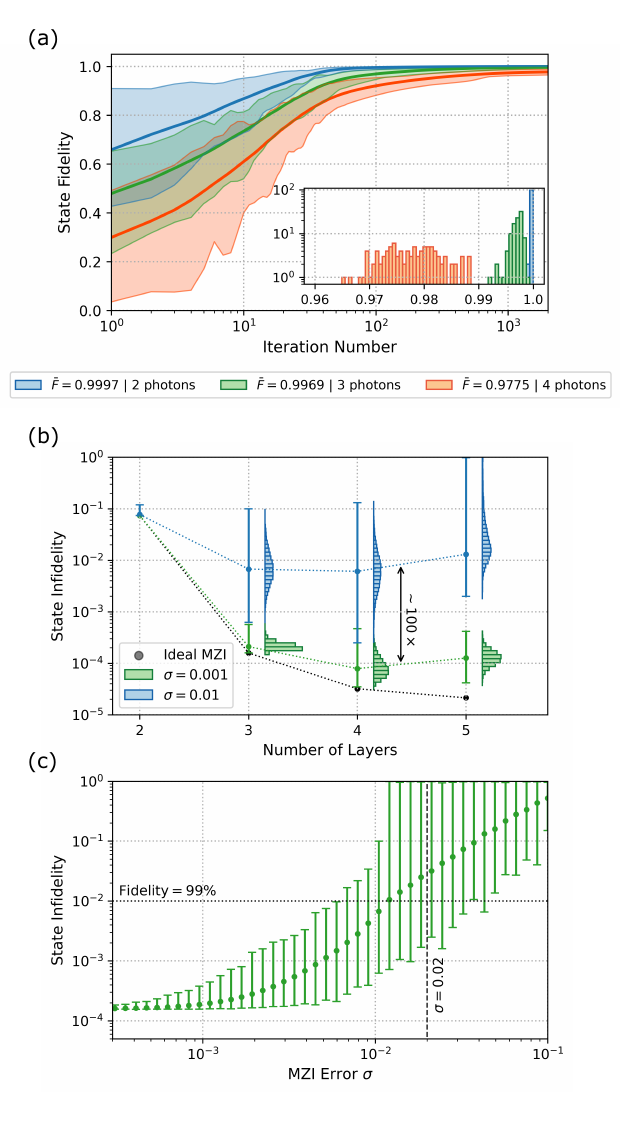}
    \caption{{\fontfamily{cmss}\selectfont\textbf{Performance of the network in learning one-to-one mapping of quantum states}. \textbf{(a)} State fidelity as a function of iteration number for a sample of 100 Haar-random multi-photon quantum states. The fidelity of the 2, 3, and 4-photon states approaches unity within 2000 iterations. The inset illustrates the distribution of learned state fidelities. Increasing the number of photons increases the difficulty in learning the target state, resulting in a decrease in the mean fidelity $\Bar{F}$. \textbf{(b)} State infidelity as a function of the depth of the network for the 4-photon $N00N$ state under the influence of component imperfections. Increasing the depth of the network increases the fidelity, denoted by the black line. The green and blue distributions correspond to the distribution of state infidelity when the beam-splitter error $\sigma$ is 0.001 and 0.01 respectively. \textbf{(c)} Distribution of state infidelity as a function of beam-splitter error $\sigma$ in the Mach-Zehnder Interferometer (MZI). Increasing the error $\sigma$ increases the mean infidelity, as well as the distribution of state fidelities.}}
    \label{fig:Haar_N00N}
\end{figure}

To demonstrate the generalizability and efficiency of the training routine, we consider the problem of preparing multimode multi-photon states. Given an initial state $\ket{\psi_{\mathrm{in}}}$, the network is trained to find the unitary operation that transforms $\ket{\psi_{\mathrm{in}}}$ into a selected target state $\ket{\psi_{\mathrm{target}}}$. We consider two separate tasks - firstly the preparation of a sample of Haar-random multi-photon states, to show that the network is able to prepare arbitrary multimode multi-photon states. Secondly, we consider the specific task of preparing a 4-photon $N00N$ state, to analyze the impact of component imperfections on the performance of the network. 

For the task of state preparation, the parameters of the neural network are optimized to maximize the overlap between a given input state and a selected target state. The fidelity of the learned state is evaluated as: 
\begin{equation}
    \mathcal{F}_{\mathrm{state}} = |\langle \psi_{\mathrm{target}} | U_{\mathrm{NN}} | \psi_{\mathrm{in}} \rangle |
    \label{eq:fid_state}
\end{equation}
where $U_{\mathrm{NN}}$ is the transformation implemented by the neural network. To find the optimal parameter set, the loss function $\mathcal{L} = (1 - \mathcal{F}_{\mathrm{state}})^{2}$ is minimized using the gradient descent technique described above. 

First, we consider the task of preparing a sample of Haar-random multi-photon states. Starting with a unitary $U$ sampled from the Haar measure~\cite{tung2003group, mezzadri2006generate}, a Haar-random quantum state $\ket{\psi_{\mathrm{Haar}}}$ is generated through the unitary evolution of a fiducial initial state $\ket{\psi_{0}}$, such that $\ket{\psi_{\mathrm{Haar}}} = U\ket{\psi_{0}}$. These Haar-random states are uniformly distributed over the space of multi-photon qudit states~\cite{meckes2019random}. Successfully preparing a sufficiently large sample of these states would empirically indicate that the network can learn mappings to prepare the entire group of multi-photon states.

A network that is 4 layers deep is trained to prepare a set of 100 Haar-random $N$-photon states in a 4-mode network. The initial state that is input into the network are single-photon Fock states that populate the first $N$ modes of the network. The state fidelity $\mathcal{F}_{\mathrm{state}}$ as a function of the number of iterations is shown in fig.~\ref{fig:Haar_N00N}(a). The blue, green and orange lines plot the average fidelity over the 100 samples for the 2, 3, and 4 photon states respectively. The shaded regions indicate the upper and lower bounds of the state fidelity over all the samples for each multi-photon state. The inset shows the distribution of the final fidelity for the 100 runs of each multi-photon state after 2000 iterations. Increasing the number of photons increases the size of the Hilbert space, and correspondingly the difficulty in training to high fidelities. This is seen in the final average fidelity, which drops from $> 99.9\%$ to $\sim 97.8\%$ as the number of photons increases from 2 to 4. We expect that fine-tuning the hyper-parameters, or annealing the learning rate further will improve the performance of the network in preparing larger photon-number states.

Next, we consider the preparation of $N00N$ states, which are many-body entangled states that are extremely sensitive to noise and loss. Here, we analyze the performance of the network under the influence of component imperfections in preparing a 4-photon $N00N$ state. Component imperfections due to fabrication process variations introduce perturbations in the splitting ratio of beam-splitters in the Mach-Zehnder Interferometers. Imperfect splitting within beam-splitter meshes introduce errors into the programmed unitary matrix~\cite{bandyopadhyay2021hardware,vadlamani2023transferable, hamerly2022asymptotically}, and thereby give rise to infidelity in the preparation of any target state. Deviations of the constituent beam-splitters from the 50:50 splitting ratio is denoted by angles $(\alpha, \beta)$ are assumed to be distributed as independent Gaussians $\mathcal{N}(0, \sigma)$.

The state infidelity as a function of the number of layers for ideal and faulty circuits is plotted in fig.~\ref{fig:Haar_N00N}(b). In the case of ideal circuits, the fidelity improves upon increasing the depth of the network, as indicated by the black line. A network that is 3 layers deep is sufficient to prepare the 4-photon $N00N$ state to a fidelity $> 99.9\%$.  To analyze the impact of faulty components on the performance of the network, we sample 10000 splitter errors $(\alpha, \beta)$  The green and blue histograms indicate distributions of state infidelities for small splitter errors ($\sigma = 0.001$) and larger, more practical splitter errors ($\sigma = 0.01$). At a depth of 4 layers, circuits with large errors already perform  $\sim 100\times$ worse than near-perfect circuits. Increasing the depth beyond 4 layers starts increasing the state infidelity, because of the vanishingly small improvements in fidelity and accumulation of a larger number of cascaded errors. The state infidelity as a function of the splitter error $\sigma$ is plotted in fig.~\ref{fig:Haar_N00N}(c) for a network that is 3 layers deep. The error bars illustrate the best and worst-case infidelities for a given beam-splitter error $\sigma$. Increasing the splitter errors corresponds to an increase in the median state infidelity as well as the upper bound of the distribution of infidelities. At $\sigma = 0.02$, which is the typical tolerance for wafer-scale process variations~\cite{prinzen2013fabrication}, the median fidelity is below $99\%$, with the worst case fidelity approaching $\sim 10\%$. A number of reconfiguration techniques have been proposed to correct for coherent errors in the interferometer platform, including global optimization~\cite{burgwal2017using, pai2019matrix, mower2015high, lopez2019programmable, lopez2020auto, perez2020multipurpose}, local correction~\cite{bandyopadhyay2021hardware, kumar2021mitigating}, and self-configuration~\cite{miller2017setting, hamerly2022accurate, hamerly2019large, pai2020parallel}.

\subsection{Universal Operations on Encoded Bases}

\begin{figure}
    \centering
    \includegraphics[width = \columnwidth]{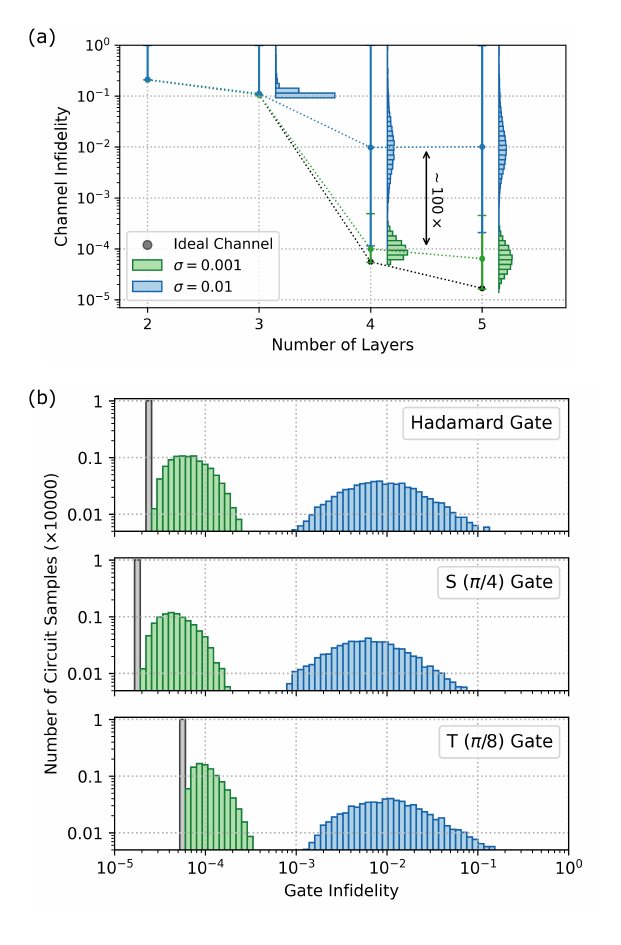}
    \caption{{\fontfamily{cmss}\selectfont\textbf{Performance of the network in learning the encoding channel and single qubit gates for the two-mode 5-photon $\chi^{(2)}$ binomial code.} \textbf{(a)} Encoding channel infidelity as a function of the depth of the network in a two-mode network. The black line indicates the infidelity of the ideally trained channel. The green and blue histograms illustrate the distribution of infidelities when splitter errors are $\sigma = 0.001$ and $\sigma = 0.01$ respectively. \textbf{(b)} Gate infidelities of single qubit logical gates (Hadamard, $S$ and $T$ phase gates) trained by a 4-layer network. The black bar indicates that an ideal gate can be trained to a fidelity $> 99.99\%$. The network performance is severely effected by component imperfections, with the worst case fidelity approaching $\sim 10\%$.}}
    \label{fig:BC_1}
\end{figure}

The versatility of the neural network architecture can also be used to realize universal transformations on encoded quantum information by implementing arbitrary quantum channels. In other words, the neural network is trained to find a completely positive trace preserving linear map $\mathcal{E}$ operating on encoded quantum information. To prepare quantum channels $\mathcal{E}$, the parameters of the network are optimized to maximize the overlap between the transformation implemented by the network $U_{\mathrm{NN}}$ and the target channel. The fidelity of the learned channel is evaluated as the average fidelity over the Haar measure~\cite{nielsen2002simple} as:
\begin{align}
    \mathcal{F}_{\mathrm{av}} (U_{\mathrm{NN}}, \mathcal{E}) & = \int \d\psi \bra{\psi}U_{\mathrm{NN}}^{\dagger}\mathcal{E} \left( \ket{\psi} \bra{\psi} \right) \ket{\psi} \nonumber \\ 
    & = \frac{d\mathcal{F} \left( \mathcal{E}, U_{\mathrm{NN}} \right) + 1}{d + 1} = \frac{\mathrm{Tr} \left( U_{\mathrm{NN}}^{\dagger} \mathcal{E} \right)+ 1}{d + 1}
    \label{eq:fid_channel}
\end{align}
where $d$ is the dimension of the multi-photon quantum system. The loss function, defined as $\mathcal{L} = (1 - \mathcal{F}_{\mathrm{av}}(U_{\mathrm{NN}}, \mathcal{E}))^{2}$ is minimized using the gradient descent technique described above to maximize the average channel fidelity $\mathcal{F}_{\mathrm{av}}$ to 1. 

The universality of the neural network architecture allows us to provide additional robustness against dominant photon loss and phase errors by performing bosonic encoding. Here, we demonstrate that the network is able to encode a bosonic error-correcting code, and prepare a universal gate set in this encoded basis. To showcase this universality, we consider an encoding that \textit{cannot} be prepared using linear optics and a $\chi^{(2)}$ optical nonlinearity. This is the two-mode $\chi^{(2)}$ binomial code proposed by Niu \textit{et al.}~\cite{niu2018hardware}, whose logical basis states are: 
\begin{align}
    \ket{\tilde{0}} & = \frac{1}{2^{N - 1}} \sum_{j=0}^{N-1}\sqrt{\binom{2N-1}{2j}}~\ket{2j, 2N-1-2j} \nonumber \\ 
    \ket{\tilde{1}} & = \frac{1}{2^{N - 1}} \sum_{j=0}^{N-1}\sqrt{\binom{2N-1}{2j+1}}~\ket{2j+1,  2(N-1-j)}
    \label{eq:BC_def}
\end{align}
where $\ket{n_{\mathrm{a}}, n_{\mathrm{b}}}$ denotes a two-mode code state with $n_{\mathrm{a}}$ and $n_{\mathrm{b}}$ photons in modes $\hat{a}$ and $\hat{b}$ respectively. The basis states of the two-mode $\chi^{(2)}$ binomial code in eqn.~\eqref{eq:BC_def} do not lie in an irreducible subspace of the $\chi^{(2)}$ Hamiltonian, and therefore cannot be prepared using just linear optics and $\chi^{(2)}$ nonlinear processes. The network architecture is not constrained by any symmetry among its modes, and should therefore be able to generate the complete Lie algebra for all qudit subspaces~\cite{krastanov2015universal, jacobs2007engineering, niu2018qudit}. This architecture is therefore ideally suited to encode and perform logical operations on the bosonic code. 

The $(2N - 1)$ photon two-mode $\chi^{(2)}$ binomial code offers an attractive option for encoding logical states because of its ability to correct photon loss and gain errors, upto $N$ photons, $N^{\mathrm{th}}$ order dephasing, and amplitude damping error. Furthermore, to protect the logical qubits up to $N$ photon losses, only $(2N-1)$ input photons are required, implying that it has a constant code rate~\cite{niu2018hardware}.

\begin{figure}
    \centering
    \includegraphics[width = \columnwidth]{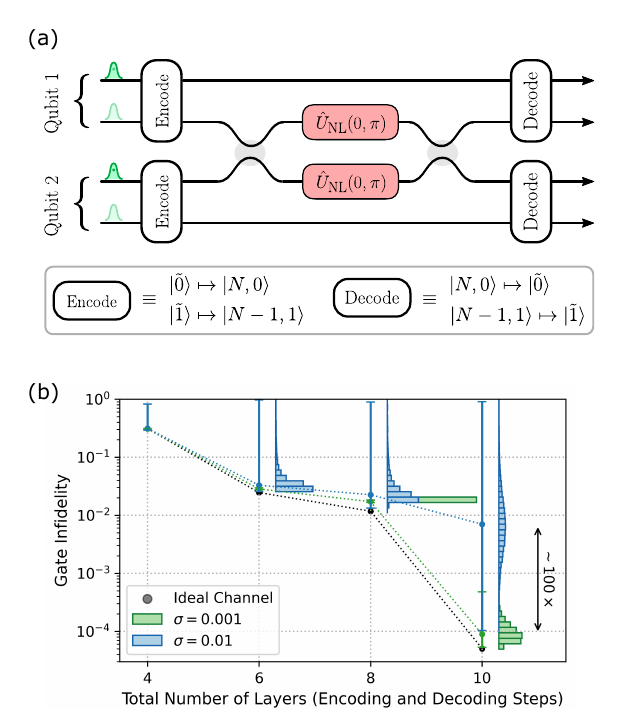}
    \caption{{\fontfamily{cmss}\selectfont\textbf{Logical controlled phase gate schematic and performance. }\textbf{(a)} Circuit implementation of the logical controlled phase gate with an encoding and decoding step that rotates the logical states into the fock basis and vice-versa. \textbf{(b)} Gate fidelity of the logical controlled phase gate illustrated in panel (a) on the 5-photon two-mode binomial code. A network that is 10 layers deep (5 layers in the encoding and decoding steps each) is able to perform the transformation to a fidelity $> 99.99\%$. The green and blue histograms indicate the distribution of infidelities when splitter errors are $\sigma = 0.001$ and $\sigma = 0.01$ respectively, with worst-case fidelities on the order of $\sim 10\%$. }}
    \label{fig:BC_2}
\end{figure}

To encode the $\chi^{(2)}$ binomial code, the network is optimized to map $(2N - 1)$ photon Fock states in each mode into the logical basis states. Fig.~\ref{fig:BC_1}(a) plots the infidelity of encoding as a function of the depth of the network to encode the 5-photon code in a two-mode network. We see that 4 layers is sufficient to  train the channel to a fidelity $> 99.99\%$. The green and blue histograms indicate the channel infidelity under the influence of component imperfections when splitter errors are ($\sigma = 0.001$) and ($\sigma = 0.01$) respectively. In this case, the splitter errors severely degrades the channel fidelity, since maintaining the relative phase among the output modes is necessary to achieve a high fidelity. Splitter errors on the order of $1\%$ drastically reduces the channel fidelity, with the worst case fidelity reaching $< 10\%$, irrespective of the depth of the network.

In order to prove that the network can perform arbitrary quantum operations, it is sufficient to show that a universal logical gate set can be constructed. We consider the generators of the logical Clifford group, namely the Hadamard gate, the $S$ ($\pi/4$) phase gate and the controlled phase gate. Along with these gates, the $T$ ($\pi/8$) phase gate completes the gate set required to construct any quantum gate in the encoded basis. Fig.~\ref{fig:BC_1}(b) shows the fidelity of the single qubit logical gates achieved by a 4-layer network. The performance of the ideal network is illustrated by the black bar, indicating that under gate fidelites greater than $99.99\%$ can be achieved. The green and blue histograms plot the distribution of gate fidelity under the influence of component imperfections. Under realistic conditions ($\sigma = 0.01$), a large portion of the sampled circuits still reaches gate fidelities above $99\%$. These networks are, however, still very sensitive to splitter errors, since maintaining the relative phase among the output logical states is essential for high-fidelity operation of the gate.

Together with the single-qubit gates, a two-qubit entangling gates in the logical basis completes the universal gate set. A schematic of the circuit used to implement the controlled-phase gate is shown in fig.~\ref{fig:BC_2}(a). This circuit uses additional encoding and decoding steps, which are implemented independently by a two-mode network. This circuit uses the same nonlinear activation $\hat{U}_{\mathrm{NL}} (0, \pi)$ described in sec.~\ref{sec:2}, but with additional encoding and decoding steps. The input to this circuit into either set of rails is the $N$-photon two-mode logical code state. First, the encoding circuit is used to rotate the logical state into the Fock basis. The encoding step is trained to perform the following transformation:
\begin{align}
    \ket{\tilde{0}} & \rightmapsto \ket{N, 0} \nonumber \\ 
    \ket{\tilde{1}} & \rightmapsto \ket{N - 1, 1}
    \label{eqn:logical_cphase}
\end{align}
where the state $\ket{n_{\mathrm{a}}, n_{\mathrm{b}}}$ denotes the a state where $n_{\mathrm{a}}$ photons populates the outer rails (uppermost and lowermost rails of the network) and $n_{\mathrm{b}}$ photons populates the inner rails of the network (the middle two rails). The logical $\ket{\tilde{0}}$ state is mapped into the $N$-photon state that populates only the outer rail, while the logical $\ket{\tilde{1}}$ state is mapped into a state with only a single photon on the inner rail. From here, it is straightforward to see that only the $\ket{\tilde{1} \tilde{1}}$ behaves similar to the $\ket{11}$ state of a traditional controlled-phase gate. Only the $\ket{\tilde{1} \tilde{1}}$ state exploits the Hong-Ou-Mandel effect and receives a $\pi$ phase shift from the nonlinear element. The state is then transformed by the decoding step, which rotates the Fock basis states back into the two-mode encoded logical state. This construction is agnostic to the encoding being used, meaning the same circuit model with updated phases can be reused for any family of bosonic codes.

The performance of the logical controlled phase gate is shown in fig.~\ref{fig:BC_2}(b). Both the encoding and decoding steps were trained independently by optimizing the fidelity metric defined in eqn.~\eqref{eq:fid_channel}. A network that is 10 layers deep (with 5 encoding layers and 5 decoding layers) is able to reach a gate fidelity $> 99.99\%$. In this case as well, component imperfections severely impact the performance of the network, where networks with larger errors perform $\sim 100\times$ worse that networks with near ideal components. The worst case fidelities reach as low as $\sim 10\%$ when $\sigma = 0.01$, irrespective of the depth of the network.

\subsection{Quantum Error Correction via Non-Demolition Measurements}

Error corrected operation of the quantum photonic neural networks is essential for the efficient performance and scaling up of such an architecture. In practice, the main decoherence channel is photon loss which imposes a bottleneck on the fidelity of the logical gate operations. Typically, error correction on bosonic codes is done by first performing a non-demolition measurement followed by a unitary correction operation conditioned on the measurement result~\cite{chuang1997bosonic, niu2018hardware, niu2018qudit, grimsmo2020quantum, albert2018performance, michael2016new, krastanov2021room}. Given photon-number parity measurements $p_{\mathrm{BC}} = \langle \hat{n}_{\mathrm{a}} + \hat{n}_{\mathrm{b}} \rangle \mathrm{mod} \left( 2N - 1 \right)$, the bosonic error-correcting codes from ref.~\cite{niu2018hardware} are able to uniquely identify the type of error that has occurred. Photon loss due to the amplitude damping channel is also shown to satisfy the Knill-Laflamme condition of our encoding.

\begin{figure}[h!]
    \centering
    \includegraphics[width = \columnwidth]{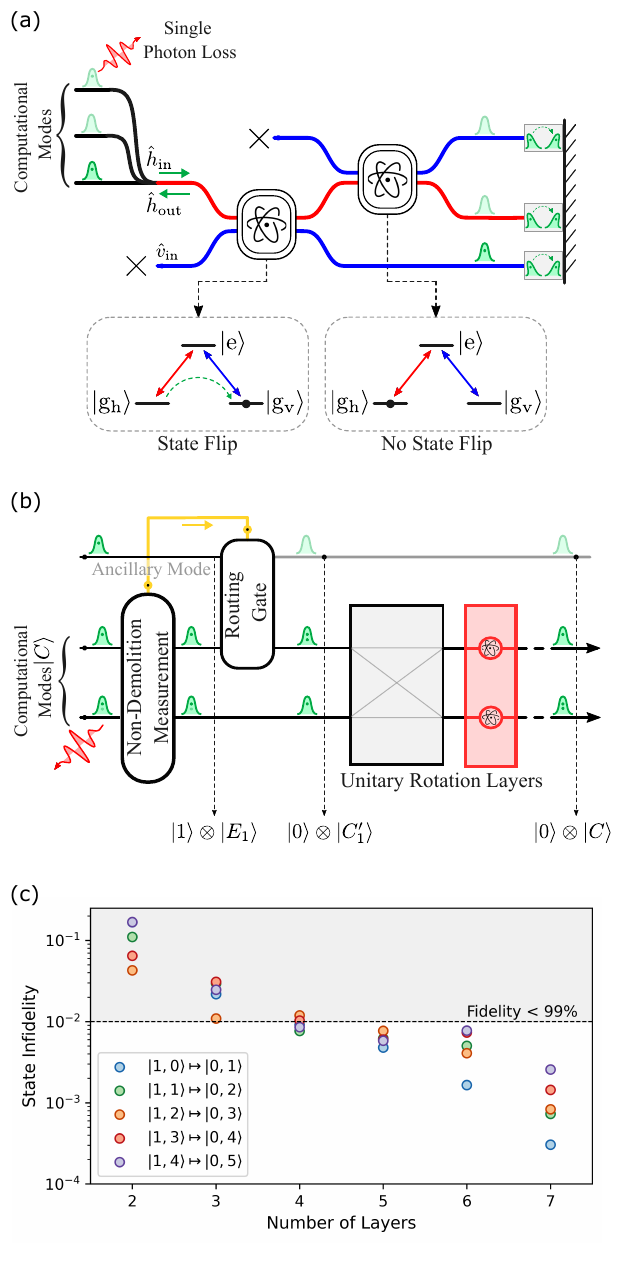}
    \caption{{\fontfamily{cmss}\selectfont\textbf{Implementation of error-correction for photon loss. }\textbf{(a)} Schematic of non-demolition measurement using the three-level $\Lambda$ atomic system. This construction allows for the detection of photon loss among two input photons from the computational modes. \textbf{(b)} Schematic of circuit to correct for single photon loss from the first mode using the non-demolition measurement, conditional routing from an ancillary mode, and unitary rotation using additional layers. \textbf{(c)} Infidelity of routing a single photon from the ancillary mode into the computational modes, when the routing gate is constructed using a separately trained network. Deeper networks result in higher fidelity for addition of the photon into all possible code states.}}
    \label{fig:EC}
\end{figure}

The strong light-matter interaction of our optical nonlinearity can be adapted to measure the total photon number non-destructively. This scheme using cascaded three-level $\Lambda$ atomic systems is illustrated in fig.~\ref{fig:EC}(a). An erroneous code state is made to sequentially interact with the cascaded atomic systems. Each interaction moves a single photon from the code state into an orthogonal mode, and deterministically flips the state of the atom. Therefore, information regarding whether an atom has interacted with a single photon is stored in the state of that atom. Measuring and resetting the state of the atoms after the photonic state has propagated through the atoms provides an error syndrome of the total photon number non-destructively. After the measurement, the photonic state that has undergone multiple subtraction steps is then reflected and time-reversed to restore the original erroneous code state. Photon-number resolving non-demolition measurements can also be performed using second or third-order optical nonlinearities~\cite{milburn1983quantum, imoto1985quantum, he2011cross, venkataraman2013phase, balybin2022quantum, yanagimoto2023quantum} or atomic systems~\cite{nogues1999seeing, chen2017quantum, niemietz2021nondestructive}. 
 
As an example, we analyze the case of performing a non-demolition measurement on a 2-photon code state. If this code state loses a single photon, the remaining photon interacts with only the first three-level $\Lambda$ system, thereby flipping its state. Measuring the state of the atoms now gives us the error syndrome $\ket{\mathrm{g}_{\mathrm{v}}}\ket{\mathrm{g}_{\mathrm{h}}}$, indicating the loss of a single photon.  In the case when there is no loss of a photon, both photons would interact with the atoms, flipping their states. This gives us the error syndrome $\ket{\mathrm{g}_{\mathrm{v}}}\ket{\mathrm{g}_{\mathrm{v}}}$, indicating there was no loss of a photon. Similarly, if both photons are lost from this state, none of the atomic systems would flip their states. This gives us error syndrome $\ket{\mathrm{g}_{\mathrm{h}}}\ket{\mathrm{g}_{\mathrm{h}}}$, indicating that both photons were lost.

Conditioned on the non-demolition measurement, the appropriate correction operations can be performed to correct the error. A schematic of this correction circuit is illustrated in fig.~\ref{fig:EC}(b). The correction circuit consists of a routing gate that is conditioned on feedback from the non-demolition measurement, and unitary rotation operations. An erroneous state prompts the routing gate to move a single photon from the ancillary mode into any one of the computational modes. The number of photons in the code space is restored, but this need not correspond to the correct original state. Therefore, additional layers of the network can be used to rotate this state into the correct state.  

To provide a more concrete example, we consider the case of a single photon loss error in a two-mode three-photon $\chi^{(2)}$ binomial code. In general, the code-space of the network can be described as $\ket{C} = \alpha\ket{\tilde{0}} + \beta\ket{\tilde{1}} = \alpha\left( \ket{0, 3} + \ket{2, 1} \right) + \beta\left( \ket{3, 0} + \ket{1, 2} \right)$. The photon loss is equally likely to occur from either mode. If there is a loss from the first mode, the state of the system becomes $\ket{E_{1}} = \alpha\ket{1, 1} + \beta\left( \ket{2, 0} + \ket{0, 2} \right)$. On the other hand, if this loss occurs from the second mode, the state of the system becomes $\ket{E_{2}} = \alpha\left( \ket{0, 2} + \ket{2, 0} \right) + \beta\ket{1, 1}$. Upon detection of this loss from the non-demolition measurement, the routing gate moves a single photon from the ancillary mode into the first mode of the code space. This transforms the error states $\ket{E_{1}}$ into $\ket{C_{1}'} = \alpha\ket{2, 1} + \beta\left( \ket{3, 0} + \ket{1, 2} \right)$, and $\ket{E_{2}}$ into $\ket{C_{2}'} =  \alpha\left( \ket{1, 2} + \ket{3, 0} \right) + \beta\ket{2, 1}$. Finally, the additional layers rotate the states $\ket{C_{1}'}$ and $\ket{C_{2}'}$ back into the original state $\ket{C}$. To summarize, the error-correction procedure performs the following transformations over the ancilla-code space: 
\begin{align}
    \ket{1} \otimes \ket{E_{1}} & \mapsto \ket{0} \otimes \ket{C_{1}'} \mapsto \ket{0} \otimes \ket{C} \nonumber \\
    \ket{1} \otimes \ket{E_{2}} & \mapsto \ket{0} \otimes \ket{C_{2}'} \mapsto \ket{0} \otimes \ket{C}
    \label{eqn:error_corr}
\end{align}

The routing gate adds a single photon into the code space conditioned on feedback from the non-demolition measurement. In other words, it transforms the two-mode state $\ket{1, n} \mapsto \ket{0, n + 1}$, where $n$ is an unknown number of photons propagating in the computational mode. One way to perform this operation is to use a photon addition scheme~\cite{gea2013photon, lund2024subtraction}. These schemes are, however, not deterministic and require pulses with specific temporal mode profiles in conjunction with weak second order nonlinearities. Therefore, we use a separately trained network tiled between the ancillary and computational modes to act as a routing gate. This network is trained to transform an input state $\ket{1, n}$ into the state $\ket{0, n + 1}$ for the 5-photon code state, where $n \in [0, 4]$. Fig.~\ref{fig:EC}(c) plots the infidelty of this gate as a function of the number of layers for all possible values of $n$. A network that is 5 layers deep reaches a fidelity greater than $99\%$ for all the required transformations. 

In practice, these operations performed by the neural network are not fault tolerant. Therefore, these simulations give us a lower bound on the hardware resources required to implement these quantum operations on an encoded basis.  Proper analysis on realistic error propagation through the optical neural network can help us mitigate these issues once combined with real-time error correction. The impact of the added complexity, and their potential for fault-tolerant error correction will be the subject of our future work.  

\section{Discussion \label{sec:4}}

Ensuring that the neural network architecture operates at high fidelity requires strong light-matter interactions under realistic hardware conditions. This necessitates that both transitions of the three-level $\Lambda$ atomic system be coupled to the cavity modes with equal cooperativity, given by:

\begin{equation}
    C = \frac{g^{2}}{2 \left( \kappa_{\mathrm{i}} + \kappa_{\mathrm{ex}} \right) \gamma}
    \label{eqn:cooperativity}
\end{equation}
where $g$ is the coupling rate of the atomic transitions to the cavity modes, $\kappa_{\mathrm{i}}$ is the intrinsic coupling rate, $\kappa_{\mathrm{ex}}$ is the extrinsic coupling rate, and $\gamma$ is the atomic emission rate into free space. Seminal work from Rosenblum \textit{et al.}~\cite{rosenblum2016extraction} shows that using $^{87}$Rb atoms coupled to a microsphere resonator, it is possible to deterministically extract a single photon from an optical pulse. The parameters achieved in their experiment are $(g, \kappa_{\mathrm{ex}}, \kappa_{\mathrm{i}}, \gamma) = (24, 40, 6.6, 3)~\mathrm{MHz}$, corresponding to a cooperativity $C \approx 8.2$. In sec.~\ref{sec:appendix_4} of the Supplementary Information, we plot the coupling strengths $g$ and $\kappa$ of recent experiments involving the integration of quantum emitters to optical cavities in different material platforms. These experiments place the achieved coupling rate $g/(2\pi)$ in the neighbourhood of $\sim 1-10~\mathrm{GHz}$, and the cavity decay rate $\kappa/(2\pi)$ in the range of $\sim 10-100~\mathrm{GHz}$. This indicates that experimental devices are already operating in the range where $g/\kappa \approx 0.1-1$, and are approaching the strong coupling regime where $g/\kappa > 1$. This suggests that the nonlinear element can be realized with near-term photonic hardware.

Along with the requirement for high cooperativity, single-mode operation of the nonlinear element also imposes the condition that the pulses are spectrally narrow. Specifically, the temporal width of the pulses is required to be longer than the cavity-enhanced decay rate ($2g^{2}/\kappa \gg \sigma$) and the line-width of the cavity ($\kappa \gg \sigma$). In sec.~\ref{sec:appendix_2} of the Supplementary Information, we evaluate the effect of temporal mode distortions on the fidelity of the nonlinear phase gate as a function of the parameters $g$ and $\kappa$. In the regime where the spectral width of the pulse $\sigma/g$ is on the order of $\sim 0.2~\kappa/g$, the fidelity of the nonlinear phase gate exceeds $99.9\%$, indicating that the nonlinear dynamics is strongly confined to the single-mode subspace. To estimate exact physical values of the pulse width required to maintain single-mode operation, we refer to the data of coupling strength $g$ and cavity decay rate $\kappa$ achieved in recent experiments involving coupling a quantum emitter to an optical cavity, plotted in sec.~\ref{sec:appendix_4} of the Supplementary Information. Assuming nominal values of $g \sim 2\pi \times 10~\mathrm{GHz}$, and $\kappa \sim 2\pi \times 50~\mathrm{GHz}$, and operating in the regime where $2g^{2}/\kappa \approx 10\sigma$, the full-width half-maximum width of the pulse can be estimated to be $\approx 1~\mathrm{ns}$. Spectrally narrow pulses with temporal widths on the order of $\sim 1~\mathrm{ns}$ can be readily generated using quantum emitters in a number of material platforms including quantum dots~\cite{rakher2011simultaneous, uppu2021quantum, kuhlmann2015transform, hennessy2007quantum} and defect centers in diamond~\cite{knall2022efficient, bradac2019quantum, babinec2010diamond} and silicon~\cite{komza2024indistinguishable, saggio2024cavity, simmons2024scalable}.

In summary, our architecture is capable of preparing and manipulating bosonic states and implementing universal operations on encoded bases. Using linear optical elements and cavity-assisted light-matter interactions, the photonic neural network is capable of performing error-corrected logical quantum computation. The optical nonlinearity that implements the element-wise activation function is a reprogrammable photon-number selective phase gate, and is based on passive light-matter interactions that does not require fast active control of cavity coupling. Moreover, the dynamics of the nonlinear element are confined to the single-mode subspace, which is essential for the construction of deterministic, high-fidelity quantum gates. We have shown its versatility in deterministically preparing a wide array of multimode multi-photon states, with applications in preparing resource states such as $N00N$ states. 

Importantly, this architecture is not constrained by any symmetries imposed by an evolution under given system Hamiltonian, such as a $\chi^{(2)}$ or $\chi^{(3)}$ process~\cite{krastanov2021room,basani2024all,niu2018hardware,niu2018qudit}. This suggests that the network is capable of performing universal operations on encoded bases, including bosonic error-correcting codes. We showcase the flexibility of the architecture in encoding and performing logical operations on the 5-photon two-mode $\chi^{(2)}$ binomial code. The simulations have assumed that the dominant source of error is coherent beam-splitter errors, and have calculated the infidelity caused by these component imperfections. The encoding step and other logical operations are also assumed to be free of photon-loss error.

Finally, by adapting the three-level $\Lambda$ atomic system, non-demolition measurements of the total photon number can be performed, which is essential for utilizing bosonic error-correcting codes. Depending on the measurement result, the appropriate unitary correction operation can be performed to correct the error. This opens up avenues for exploring alternative physical computing architectures beyond the traditional qubit basis for problem-specific and hardware-efficient ansatze. 

Advances in integrated photonics and nano-fabrication techniques has enabled the monolithic integration of a number of modules for on-chip quantum photonics~\cite{kim2020hybrid}. Our estimates indicate that the individual components of the processor can be constructed using present-day photonic hardware. While integrating all the components of the neural network onto a single platform would be experimentally challenging~\cite{kim2017hybrid}, investments in such a device would open doors to large-scale, fault-tolerant photonic quantum computation, simulation, sensing and communication.\\

\textbf{Data Availability:} All requests for code and data should be made to J.R.B at \url{jasvith@umd.edu}. \\

\textbf{Code Availability:} Source code for the simulations and the accompanying tutorials can be found in the \texttt{CasOptAx} package at \url{https://github.com/JasvithBasani/CasOptAx}

\section*{Acknowledgements}

The authors would like to thank the \texttt{JAX} open source community. Simulations of the three-level atomic system were performed using the \texttt{DIFFRAX} package ~\cite{kidger2021on}. The simulations presented in this paper were performed on the Zaratan supercomputing cluster. The authors acknowledge the University of Maryland supercomputing resources (\url{http://hpcc.umd.edu}) made available for conducting the research reported in this paper.

\clearpage

\bibliography{bib_files/nonlinear_refs, bib_files/onn_mesh_refs, bib_files/qc_refs, bib_files/qec_refs, bib_files/math_refs, bib_files/hardware_refs, bib_files/qml_refs}

\clearpage

\appendix
\onecolumngrid
\section*{Supplementary Information - Scalable Quantum Photonic Neural Network Processor via Cavity-Assisted Interactions}
\renewcommand{\theequation}{A.\arabic{equation}}
\setcounter{equation}{0}

\subsection{Nonlinear Phase Gate \label{sec:appendix_1}}

In sec.~\ref{sec:2} of the manuscript, we introduced the nonlinear activation function based on coherent photon subtraction followed by addition, using the three-level atomic system with energy levels in a $\Lambda$ configuration. To describe this system, we follow the method introduced in~\cite{gea2013photon}. The Hamiltonian of light interacting with a single three-level atom in a cavity~\cite{gea2013space} is:

\begin{equation}
    \hat{H} = -i\hbar g\sqrt{\frac{\kappa}{\pi}} \int \frac{1}{\kappa - i\omega} \Big( \ket{\mathrm{e}}\bra{\mathrm{g}_{\mathrm{h}}} \hat{a}_{\omega}  
    + \ket{\mathrm{e}} \bra{\mathrm{g}_{\mathrm{v}}} \hat{b}_{\omega} \Big)  e^{-i(\omega + \delta)t} \d\omega + \mathrm{H.C.}
\end{equation}
where $\hat{a}_{\omega} \left( \hat{a}^{\dagger}_{\omega} \right)$ and $\hat{b}_{\omega} \left( \hat{b}^{\dagger}_{\omega} \right)$ are the bosonic annihilation (creation) operators for modes $\hat{h}_{\mathrm{in}}$ and $\hat{v}_{\mathrm{in}}$ respectively. Here, $\omega$ is measured from the resonance frequency of the atom, and $\delta$ is the detuning of the cavity resonance from the atomic transition, which is assumed to be 0. The parameter $g$ denotes the coupling strength between the cavity mode and the atomic transition, and $\kappa$ denotes the cavity decay rate. The state of the system at any given time can be written as $\ket{\psi_{\mathrm{h}}}\ket{\mathrm{g}_{\mathrm{h}}} + \ket{\psi_{\mathrm{v}}}\ket{\mathrm{g}_{\mathrm{v}}} + \ket{\psi_{\mathrm{e}}}\ket{\mathrm{e}}$. The coupled equations of motion are:

\begin{equation}
    \frac{\d\ket{\psi_{\mathrm{h}}}}{\d t} = g\sqrt{\frac{\kappa}{\pi}} \int \frac{1}{\kappa + i\omega} \hat{a}^{\dagger}_{\omega} \ket{\psi_{\mathrm{e}}} e^{i\omega t} \d\omega
    \label{eq:app_langevin_1}
\end{equation}
\begin{equation}
    \frac{\d\ket{\psi_{\mathrm{v}}}}{\d t} = g\sqrt{\frac{\kappa}{\pi}} \int \frac{1}{\kappa + i\omega} \hat{b}^{\dagger}_{\omega} \ket{\psi_{\mathrm{e}}} e^{i\omega t} \d\omega
    \label{eq:app_langevin_2}
\end{equation}
\begin{equation}
    \frac{\d\ket{\psi_{\mathrm{e}}}}{\d t} = -g\sqrt{\frac{\kappa}{\pi}} \int \frac{1}{\kappa - i\omega} \left( \hat{a}_{\omega} \ket{\psi_{\mathrm{h}}} + \hat{b}_{\omega}\ket{\psi_{\mathrm{v}}} \right) e^{-i\omega t} \d\omega
    \label{eq:app_langevin_3}
\end{equation}
Integrating eqns.~\eqref{eq:app_langevin_1},~\eqref{eq:app_langevin_2} and plugging the solution into eqn.~\eqref{eq:app_langevin_3}
, we get:

\begin{align}
    \frac{\d\ket{\psi_{\mathrm{e}}}}{\d t} = & - \frac{g^{2}\kappa}{\pi} \int \d\omega \int \d\omega' \int_{0}^{t} \d t' \left( \frac{1}{\kappa - i\omega} \frac{1}{\kappa + i\omega'} \hat{a}_{\omega} \hat{a}_{\omega'}^{\dagger} \ket{\psi_{\mathrm{e}}(t')} \right) e^{-i(\omega t - \omega' t')} \nonumber \\ 
    & - \frac{g^{2}\kappa}{\pi} \int \d\omega \int \d\omega' \int_{0}^{t} \d t' \left( \frac{1}{\kappa - i\omega} \frac{1}{\kappa + i\omega'} \hat{b}_{\omega} \hat{b}_{\omega'}^{\dagger} \ket{\psi_{\mathrm{e}}(t')} \right) e^{-i(\omega t - \omega' t')} \nonumber \\ 
    & -g\sqrt{\frac{\kappa}{\pi}} \int \d\omega \frac{1}{\kappa - i\omega} \left( \hat{a}_{\omega}\ket{\psi_{\mathrm{h}}(0)} + \hat{b}_{\omega}\ket{\psi_{\mathrm{v}}(0)} \right) e^{-i\omega t}
    \label{eq:app_langevin_3_new}
\end{align}
In the main text of the manuscript, we assumed that the pulses were sufficiently long, to not introduce wavefunction distortions. In this limit, the pulse duration $T = 1/\sigma$ is faster than the cavity decay rate $\kappa$, where $T \gg 1/\kappa$ or $\kappa \gg \sigma$. Operating in this adiabatic limit, $\omega$ can be neglected on the scale of $\kappa$,~\eqref{eq:app_langevin_3_new} can be written as:

\begin{equation}
    \frac{\d\ket{\psi_{\mathrm{e}}}}{\d t} = -\frac{2g^{2}}{\kappa} \int_{0}^{t} \d t' \hat{a}_{t} \hat{a}_{t'}^{\dagger} \ket{\psi_{\mathrm{e}}(t')}
    -\frac{2g^{2}}{\kappa} \int_{0}^{t} \d t' \hat{b}_{t} \hat{b}_{t'}^{\dagger} \ket{\psi_{\mathrm{e}}(t')} - g\sqrt{\frac{2}{\kappa}} \left( \hat{a}_{t}\ket{\psi_{\mathrm{h}}(0)} + \hat{b}_{t}\ket{\psi_{\mathrm{v}}(0)} \right)
    \label{eq:app_langevin_3_simple}
\end{equation}
where $\hat{a}_{t} = \sqrt{1/2\pi}\int \hat{a}_{\omega} e^{-i\omega t} \d\omega$, which satisfies the standard commutation relation $[\hat{a}(t), \hat{a}^{\dagger}(t')] = \delta(t - t')$. This form lets us define the cavity-enhanced atomic decay rate $\Gamma = 2g^{2}/\kappa$. 

\subsubsection*{Step 1: Photon Subtraction}

For a given waveguide implementing the activation function, we assume the input is a horizontally polarized $N$-photon fock state, given by:
\begin{equation}
    \ket{\psi_{\mathrm{h}}(0)} = \frac{1}{\sqrt{N!}} \left( \int \xi(\omega) \hat{a}^{\dagger}_{\omega} \d\omega \right)^{N} \ket{0}
    \label{eq:app_psi_h_in}
\end{equation}
where $\xi(\omega)$ is the spectral profile of the pulse, such that $\int |\xi(\omega)|^{2} \d\omega = 1$. Below, we denote the time-domain representation of the pulse as $\xi(t)$, obtained through the Fourier transform of $\xi(\omega)$. We assume the atom is initially in the ground state $\ket{\mathrm{g}_{\mathrm{h}}}$. Since there are photons incident on the three-level atom from mode $\hat{v}_{\mathrm{in}}$ ($\ket{\psi_{\mathrm{v}} (0)} = 0$), we can reduce eqn.~\eqref{eq:app_langevin_3_simple} to:

\begin{equation}
    \frac{\d\ket{\psi_{\mathrm{e}}}}{\d t} = - \Gamma\ket{\psi_{\mathrm{e}}} - \Gamma \int_{0}^{t} \d t' \hat{a}_{t} \hat{a}_{t'}^{\dagger} \ket{\psi_{\mathrm{e}} (t')} - \sqrt{\Gamma} \hat{a}_{t} \ket{\psi_{\mathrm{h}}(0)}
\end{equation}
Along with the requirement that $T \gg 1/\kappa$, we also operate in the limit where the pulse is longer than $1/\Gamma$ or $\Gamma \gg \sigma$. This implies that we require $2g^{2}T \gg \kappa \implies 2g^{2}/\kappa \gg \sigma$. In this limit, under the slowly-varying pulse approximation, $\d \ket{\psi_{\mathrm{e}}}/\d t$ can be adiabatically eliminated and approximated to 0. This gives us the following simplified form:

\begin{equation}
    \ket{\psi_{\mathrm{e}}} \approx - \int_{0}^{t} \d t' \hat{a}_{t}\hat{a}^{\dagger}_{t'} \ket{\psi_{\mathrm{e}}(t')} - \frac{1}{\sqrt{\Gamma}} \hat{a}_{t} \ket{\psi_{\mathrm{h}}(0)}
    \label{eq:app_simple_psi_e}
\end{equation}
Using the commutation relation $[\hat{a}_{t}, \hat{a}^{\dagger}_{t'}] = \delta(t - t')$ and normal ordering, the right-hand side of eqn.~\eqref{eq:app_simple_psi_e} can be substituted into itself recursively to give us:

\begin{equation}
    \ket{\psi_{\mathrm{e}}} \approx - \frac{1}{\sqrt{\Gamma}} \left[ \hat{a}_{t} \ket{\psi_{\mathrm{h}}(0)} - \int_{0}^{t} \d t' \hat{a}^{\dagger}_{t'} \hat{a}_{t} \hat{a}_{t'} \ket{\psi_{\mathrm{h}}(0)} + \int_{0}^{t} \d t' \int_{0}^{t'} \d t'' \hat{a}^{\dagger}_{t''} \hat{a}^{\dagger}_{t'} \hat{a}_{t} \hat{a}_{t'} \hat{a}_{t''} \ket{\psi_{\mathrm{h}}(0)} - \hdots  \right] 
\end{equation}
Clearly, this series terminates after $N$ terms for an $N$-photon fock state input. Using the input state $\ket{\psi_{\mathrm{h}}(0)}$ from eqn.~\eqref{eq:app_psi_h_in}, by repeated integration by parts and using the commutation relations, the series can be simplified to:

\begin{equation}
    \ket{\psi_{\mathrm{e}}} = \frac{\xi(t)}{\sqrt{\Gamma}} \sum_{n = 1}^{N} \frac{(-1)^{n}}{(n - 1)!} \sqrt{\frac{N!}{\left( N - n \right)!}} \left( \int_{0}^{t}~\d t' \xi(t') \hat{a}^{\dagger}_{t'} \right)^{n - 1} \ket{N - n} = -\frac{\xi(t)}{\sqrt{\Gamma}} \sqrt{\frac{N}{(N - 1)!}} ~ \left( \int_{t}^{\infty} \d t' \xi(t') \hat{a}^{\dagger}_{t'} \right)^{N - 1} \ket{0}
    \label{eq:app_psi_e_sol}
\end{equation}
This can now be used to calculate the output state of the atom when it decays back into the ground state. Clearly, eqns.~\eqref{eq:app_langevin_1},~\eqref{eq:app_langevin_2} can be re-written in the time domain under the above mentioned adiabatic limit as:

\begin{equation}
    \frac{\d \ket{\psi_{\mathrm{h}}}}{\d t} = \sqrt{\Gamma} \hat{a}^{\dagger}_{t} \ket{\psi_{\mathrm{e}}}
    \label{eq:app_langevin_1_time}
\end{equation}
\begin{equation}
    \frac{\d \ket{\psi_{\mathrm{v}}}}{\d t} = \sqrt{\Gamma} \hat{b}^{\dagger}_{t} \ket{\psi_{\mathrm{e}}}
    \label{eq:app_langevin_2_time}
\end{equation}
By substituting eqn.~\eqref{eq:app_psi_e_sol} into the above equations, it is clear to see by inspection that $\ket{\psi_{\mathrm{h}} (t \to \infty)} = 0$. Similarly, we can solve for $\ket{\psi_{\mathrm{v}} (t)}$ to get: 
\begin{equation}
    \ket{\psi_{\mathrm{v}} (t)} = -\sqrt{\frac{N}{(N - 1)!}} \left( \int_{-\infty}^{t} \d t' \xi(t') \hat{b}^{\dagger}_{t'}   \right) \times \left( \int_{t'}^{\infty} \d t'' \xi(t'') \hat{a}^{\dagger}_{t''} \right)^{N - 1} \ket{0}
    \label{eq:app_subtracted_state_t}
\end{equation}
At a time sufficiently long after the interaction, i.e., as $t \to \infty$, the subtracted state is given by the following equation: 
\begin{equation}
    \ket{\psi_{\mathrm{v}} (t \to \infty)} = -\sqrt{\frac{N}{(N - 1)!}} \left( \int_{-\infty}^{\infty} \d t' \xi(t') \hat{b}^{\dagger}_{t'}   \right) \times \left( \int_{t'}^{\infty} \d t'' \xi(t'') \hat{a}^{\dagger}_{t''} \right)^{N - 1} \ket{0}
    \label{eq:app_subtracted_state_inf}
\end{equation}
This state indicates that the subtracted state, given by eqn.~\eqref{eq:app_subtracted_state_inf} is temporally entangled. The mode $\hat{h}'$ now contains $(N - 1)$ excitations, while the mode $\hat{v}'$ has a single excitation. Each mode, however, remains confined to the single-mode subspace, i.e., the input and output states have temporal profiles $\xi(t)$. During this process, the state of the three-level atom has also deterministically transitioned to the ground state $\ket{\mathrm{g}_{\mathrm{v}}}$.

\subsubsection*{Step 2: Time-Reversal By Phase-Conjugating Mirror}

In order to deterministically perform photon addition, the process of photon subtraction will have to be time-reversed. Unlike photon subtraction, an unentangled state cannot be added deterministically, as proven in ref.~\cite{gea2013photon}. However, time-reversing the process of photon subtraction can be used to perform deterministic photon addition, which we demonstrate below. If the pulse $\xi(t)$ is symmetric in time, i.e., $\xi(t) = \xi(-t)$, the time-reversed state is given by:
\begin{equation}
    \ket{\psi_{\mathrm{v}} (-t, -t', -t'')} = -\sqrt{\frac{N}{(N - 1)!}} \left( \int_{t}^{\infty} \d t' \xi(t') \hat{b}^{\dagger}_{t'}   \right) \times \left( \int_{-\infty}^{t'} \d t'' \xi(t'') \hat{a}^{\dagger}_{t''} \right)^{N - 1} \ket{0}
    \label{eq:app_time_rev}
\end{equation}

\subsubsection*{Step 3: Coherent Photon Addition}

In order to achieve perfect photon addition, the process of photon subtraction will have to be time-reversed. Here, we show that by using a time-reversed version of the subtracted state, the photons can be deterministically added into an $N$-photon fock state. We consider the time-reversed state in eqn.~\eqref{eq:app_time_rev} as input to a three-level atom initialized in the $\ket{\mathrm{g}_{\mathrm{v}}}$ state. The input state to be added is therefore:
\begin{equation}
    \ket{\psi_{\mathrm{in}}} = -\sqrt{\frac{N}{(N - 1)!}} e^{i\varphi_{1}} \left( \int_{t}^{\infty} \d t' \xi(t') \hat{b}^{\dagger}_{t'}   \right) \times e^{i(N - 1)\varphi_{2}} \left( \int_{-\infty}^{t'} \d t'' \xi(t'') \hat{a}^{\dagger}_{t''} \right)^{N - 1} \ket{0}
    \label{eq:app_add_input}
\end{equation}
Without loss of generality, we assume that $\varphi_{1} = \varphi_{2} = 0$. It is straightforward to see how these phase shifts will be carried forward through the photon addition process. In this case, eqn.~\eqref{eq:app_langevin_3_simple} reduces to:
\begin{equation}
    \frac{\d\ket{\psi_{\mathrm{e}}}}{\d t} = - \Gamma\ket{\psi_{\mathrm{e}}} - \Gamma \int_{0}^{t} \d t' \hat{a}_{t} \hat{a}_{t'}^{\dagger} \ket{\psi_{\mathrm{e}} (t')} - \sqrt{\Gamma} \hat{b}_{t} \ket{\psi_{\mathrm{v}}(0)}
\end{equation}
Using the same method done above, under the slowly-varying pulse approximation, with the appropriate ordering of operators and by substituting the right-hand side of the equation into itself, we get the series:
\begin{equation}
    \ket{\psi_{\mathrm{e}}} \approx - \frac{1}{\sqrt{\Gamma}} \left[ \hat{b}_{t} \ket{\psi_{\mathrm{in}}(0)} - \int_{0}^{t} \d t' \hat{a}^{\dagger}_{t'} \hat{a}_{t} \hat{b}_{t'} \ket{\psi_{\mathrm{in}}(0)} + \int_{0}^{t} \d t' \int_{0}^{t'} \d t'' \hat{a}^{\dagger}_{t''} \hat{a}^{\dagger}_{t'} \hat{a}_{t} \hat{a}_{t'} \hat{b}_{t''} \ket{\psi_{\mathrm{in}}(0)} - \hdots  \right] 
\end{equation}
Similar to the above case, this series terminates after $N$ terms. Substituting the time-reversed state from eqn.~\eqref{eq:app_time_rev} here, we find get the following series expansion: 
\begin{equation}
    \ket{\psi_{\mathrm{e}}} = -\frac{1}{\sqrt{\Gamma}} \sqrt{\frac{N}{(N - 1)!}} \left[ \xi(t) \left( \int_{-\infty}^{t'} \d t'' \xi(t'') \hat{a}^{\dagger}_{t''} \right)^{N - 1} - \sqrt{N - 1}~ \xi(t) \left( \int_{0}^{t} \d t'\xi(t')\hat{a}^{\dagger}_{t'} \right) \left( \int_{-\infty}^{t'} \d t'' \xi(t'') \hat{a}^{\dagger}_{t''} \right)^{N - 2} + \hdots  \right] \ket{0}
\end{equation}
This series can be simplified into the following expressions, and by formally extending the limits of integration from $[-\infty, +\infty]$ we get:
\begin{align}
    \ket{\psi_{\mathrm{e}}} & = -\frac{\xi(t)}{\sqrt{\Gamma}} \sum_{n = 1}^{N} \frac{(-1)^{n}}{(n - 1)!} \sqrt{\frac{N}{(N - n)!}} \left( \int_{0}^{t} \d t' \xi(t')\hat{a}^{\dagger}_{t'} \right)^{n - 1} \left( \int_{-\infty}^{t'} \d t'' \xi(t'') \hat{a}^{\dagger}_{t''} \right)^{N - n} \nonumber \\ 
    & = -\frac{\xi(t)}{\sqrt{\Gamma}} \sqrt{N} \left( \int_{-\infty}^{t} \d t'\xi(t')\hat{a}^{\dagger}_{t'} \right)^{N - 1}
\end{align}
This can now be substituted into eqns.~\eqref{eq:app_langevin_1_time} and~\eqref{eq:app_langevin_2_time} and integrated from 0 to $t$. Clearly, $\ket{\psi_{\mathrm{v}} (t \to \infty)}$ goes to 0, since its derivative equals the right-hand side of eqn.~\eqref{eq:app_add_input}, and its value at $t = 0$ is the correct input state. On the other hand, the $\ket{\psi_{\mathrm{h}}}$ state can be solved for as:
\begin{equation}
    \ket{\psi_{\mathrm{h}}} = \sqrt{N} \int_{-\infty}^{t} \d t~\xi(t)\hat{a}^{\dagger}_{t'} \times \left( \int_{-\infty}^{t} \d t'\xi(t')\hat{a}^{\dagger}_{t'} \right)^{N - 1}
\end{equation}
Clearly, as $t \to \infty$, we get $\ket{\psi_{\mathrm{h}}} = \ket{N}$, i.e., an $N$-photon fock state in the mode $\hat{a}$. This implies that the atom has deterministically transitioned from the state $\ket{\mathrm{g}_{\mathrm{v}}}$ into the state $\ket{\mathrm{g}_{\mathrm{h}}}$, and thereby moved the single photon from the $\hat{b}$ mode into the $\hat{a}$ mode. This shows that a time-reversed version of the subtracted state can be deterministically added, resulting in an $N$-photon fock state in the output mode with the corresponding nonlinear phase shift.

\clearpage
\twocolumngrid

\subsection{Fidelity of the Nonlinear Sign Gate \label{sec:appendix_2}}

In this nonlinear phase gate, the main source of infidelity arises from coupling to parasitic temporal modes. This creates undesired temporal correlations that are associated with the formation of photon bound states and higher-order extended states. Unlike the correctly scattered state (indicated in eqn.~\eqref{eq:subtracted_state} of the main text), the bound states manifest as bunched photons that propagate along \textit{one} of the optical paths, and therefore acquiring the incorrect phase. In other words, the subtracted state consists of scattered and bound state components that acquire phases \textit{relative} to each other. This causes the time-reversed photon addition to be imperfect, leading to decrease in the gate fidelity. Therefore the fidelity of the nonlinear sign gate is a function of the phase acquired by these components, and can be parameterized by $(\varphi_{1}, \varphi_{2})$. 

\begin{figure}
    \centering
    \includegraphics[width = \columnwidth]{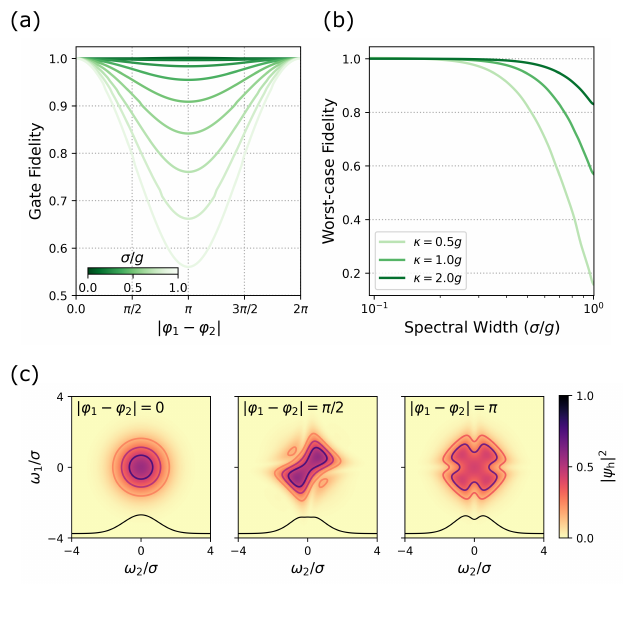}
    \caption{{\fontfamily{cmss}\selectfont\textbf{Performance metrics of the nonlinear phase gate.} \textbf{(a)} Fidelity of the sign gate as a function of the phases $|\varphi_{1} - \varphi_{2}|$ for varying spectral pulse widths $\sigma/g$ when $\kappa = 1.0g$ and $N = 2$. In the limit of spectrally narrow pulses, the contribution from the bound state vanishes, leading to high gate fidelity independent of the phases $\varphi_{1}$ and $\varphi_{2}$. \textbf{(b)} Worst-case fidelity (when $|\varphi_{1} - \varphi_{2}| = \pi$) as a function of the spectral width of the pulse for varying ratios of $g/\kappa$. As the cavity decay rate $\kappa$ increases for a fixed value of $g$, the contribution of the bound state reduces, allowing for spectrally larger pulses to achieve higher fidelities.   \textbf{(c)} Sample of spectral mode profiles after transmission of the pulse through the nonlinear sign gate with different phases $\varphi_{1}$ and $\varphi_{2}$, when $\sigma/g = 1.0$ and $\kappa = 1.0g$. When $\varphi_{1} = \varphi_{2}$, the subtracted state undergoes perfect time-reversal and results in a Gaussian mode with a global phase. As $|\varphi_{1} - \varphi_{2}|$ approaches $\pi$, the components of the bound state acquire different phases, resulting in imperfect time-reversal and distortion of the spectral mode profile. Here, as $|\varphi_{1} - \varphi_{2}|$ approaches $\pi$, the fidelity decreases from 1 to $\sim 0.55$. }}
    \label{fig:nonlinear_data}
\end{figure}

To evaluate the fidelity of the nonlinear phase gate, we numerically solve the equations of motion when an $N = 2$ photon Fock state is incident on the atom. The fidelity is calculated using the overlap integral: 
\begin{equation}
    \mathcal{F} = \int \d \omega_{1} \d \omega_{2} ~ \xi_{\mathrm{out}} (\omega_{1}, \omega_{2})~ \xi_{\mathrm{in}}^{*} (\omega_{1}, \omega_{2})
    \label{eq:overlap_integral}
\end{equation}
where $\xi(\omega)$ denotes the spectral profile of the pulses. We consider pulses with a Gaussian spectral profile, given by $\xi(\omega) = \frac{1}{\left( \pi \sigma'^{2} \right)^{\frac{1}{4}}} e^{-\omega^{2}/2\sigma'^{2}}$, where $\sigma = 2\sqrt{\mathrm{ln}(2)}\sigma'$. In fig.~\ref{fig:nonlinear_data}(a), we plot the gate fidelity as a function of the phases $|\varphi_{1} - \varphi_{2}|$ for pulses with varying spectral widths ($\sigma/g$). Spectrally narrow pulses, indicated by the darker color, achieve high gate fidelity irrespective of the phases because the contribution to the bound state is vanishingly small in the limit where $2g^{2}/\kappa \gg \sigma$. Upon increasing the spectral width of the pulses (indicated by lighter shades) the gate fidelity first decreases until it reaches a minimum at $|\varphi_{1} - \varphi_{2}| = \pi$ and then starts increasing. As $|\varphi_{1} - \varphi_{2}|$ increases from 0 to $\pi$, the relative phase among the components of the subtracted state increases, giving rise to imperfect photon addition and therefore larger gate infidelity. The worst-case fidelity is obtained at $|\varphi_{1} - \varphi_{2}| = \pi$, when the components of the subtracted state destructively interfere with each other.

These temporal mode distortions are also influenced by cavity decay rate $\kappa$. By operating in the limit where $\kappa/\sigma \gg 1$, these distortions can be adiabatically eliminated, allowing the nonlinear dynamics to be confined to the single-mode subspace. Fig.~\ref{fig:nonlinear_data}(b) plots the worst-case gate fidelity (at $|\varphi_{1} - \varphi_{2}| = \pi$) as a function of the spectral width of the pulse for various cavity decay rates. For a given cavity decay rate $\kappa$, decreasing the spectral pulse width $\sigma$ increases the fidelity that can be achieved.

Fig.~\ref{fig:nonlinear_data}(c) illustrates a a sample of the output spectral mode profiles when a two-photon Gaussian pulse of width $\sigma/g = 1.0$ is incident on the nonlinear phase gate. When $|\varphi_{1} - \varphi_{2}| = 0$, the components of the subtracted state are in-phase, and can be added perfectly. This is indicated in the leftmost panel, where there is no temporal mode distortion. As $|\varphi_{1} - \varphi_{2}|$ increases and approaches $\pi$, the phase among the components of the subtracted state results in imperfect photon addition. This imperfect photon addition distorts the temporal mode profile of the output pulse, shown in the middle and right panels.

\clearpage
\subsection{Programmable Phase Gates for Quantum Circuits \label{sec:appendix_3}}

\begin{figure}[h!]
    \centering
    \includegraphics[width = \columnwidth]{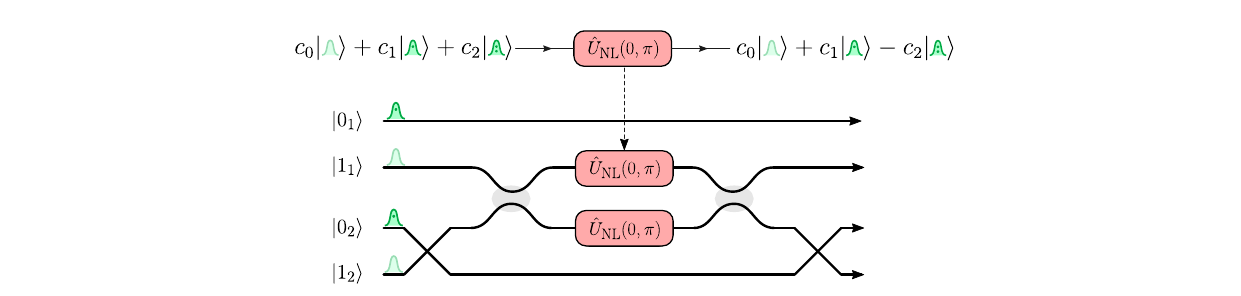}
    \caption{{\fontfamily{cmss}\selectfont\textbf{Schematic of the controlled-phase gate}. The proposed optical nonlinearity implements a nonlinear sign gate, which can be used to build a controlled-phase gate that completes a universal gate set.}}
    \label{fig:cphase_gate}
\end{figure}

The programmable nonlinearity proposed in the main text of the manuscript enables the construction of a universal gate set. It transforms the superposition state $c_{0}\ket{0} + c_{1}\ket{1} + c_{2}\ket{2}$ into the state $c_{0}\ket{0} + c_{1}e^{i\varphi_{1}}\ket{1} + c_{2}e^{i(\varphi_{1} + \varphi_{2})}\ket{2}$. When $\varphi_{1} = 0$ and $\varphi_{2} = \pi$, the transformation realizes a nonlinear sign gate $c_{0}\ket{0} + c_{1}\ket{1} - c_{2}\ket{2}$. Fig.~\ref{fig:cphase_gate} shows how this can be used to implement a controlled-phase gate. This circuit exploits the Hong-Ou-Mandel effect to ensure that two photons are incident on the nonlinear element when the qubits are in the state $\ket{11}$, implementing a $\pi$ phase shift exclusively for this state.

\begin{figure}[h!]
    \centering
    \includegraphics[width = \columnwidth]{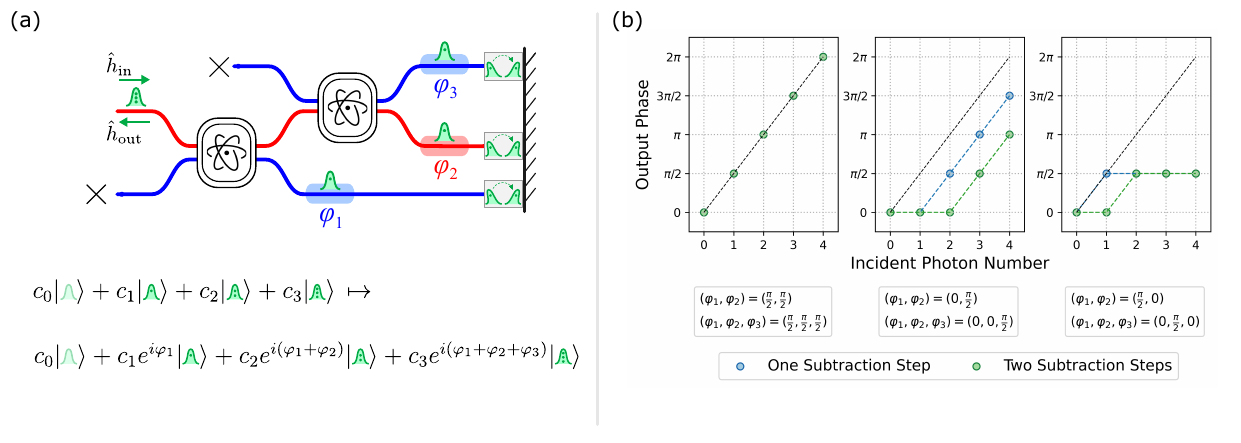}
    \caption{{\fontfamily{cmss}\selectfont\textbf{Programmable nonlinearity for photon-number selective arbitrary phase gate}. \textbf{(a)} Schematic of a nonlinear element with two cascaded $\Lambda$ atomic systems. This allows for two subtraction steps, enabling a photon-number selective phase gate for $N = 3$ incident photons. \textbf{(b)} Examples of programmable activation functions with one and two photon subtraction steps. These functions are linear when $N > 3$. The phases are selected to resemble the common Rectified Linear Unit (ReLU) and Sigmoid functions.}}
    \label{fig:snap_gate}
\end{figure}

This model of nonlinear activation is that it can be extended to $N > 2$ photons in the incident pulse. Activation functions for $N > 2$ are however, not fully tunable photon-number selective phase gates. Since only a single photon is subtracted, this model only allows sufficient degrees of freedom to select phases for pulses with up to 2 photons. If a pulse contains $N > 2$ photons, the activation function is linear in phase, scaling as $(N - 1) \times \varphi_{2}$.  To allow full programmability, i.e., realize a fully photon-number selective arbitrary phase gate, multiple three-level $\Lambda$ systems can be used to implement multiple subtraction steps. This would 'unravel' the pulse into its constituent single photons, and each of them can then acquire an independent phase shift. An example of an arbitrary photon-number selective phase gate for $N = 3$ is illustrated in fig.~\ref{fig:snap_gate}(a). Each subtraction step places a single photon into a separate mode, where it can acquire an independent phase shift. The time-reversal and second pass through the atomic systems would then 'repack' the constituent photons into the original pulse. To construct an arbitrary phase gate for $N$ photons, one would require $(N - 1)$ three-level $\Lambda$ systems. 

Fig.~\ref{fig:snap_gate}(b), illustrates some examples of nonlinear activation functions that can be programmed. The blue and green curves depict the functions that can be implemented when there are one and two subtraction steps respectively. Increasing the number of subtraction steps by adding more atomic systems allows a broader class of functions to be expressed.

\onecolumngrid

\clearpage

\subsection{Experimental Progress \label{sec:appendix_4}}

\twocolumngrid

\begin{figure}
    \centering
    \includegraphics[width = \columnwidth]{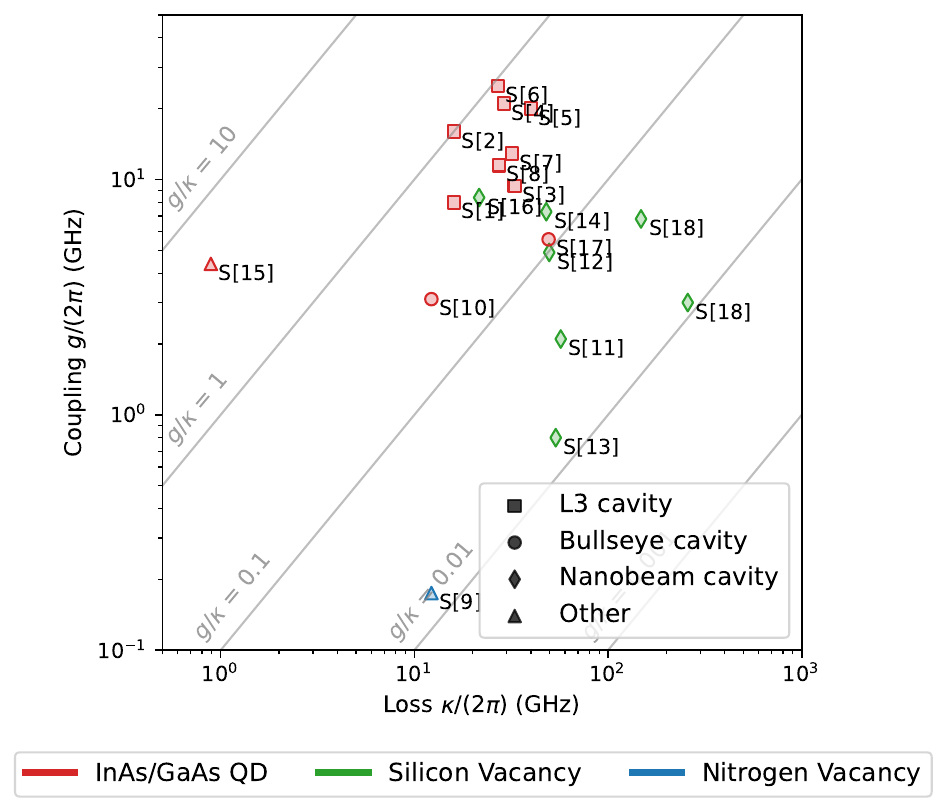}
    \caption{{\fontfamily{cmss}\selectfont\textbf{Recent experiments requiring quantum emitters coupled to optical cavities}. Coupling of the atomic transition ($g$) and the cavity decay rate ($\kappa$) for different types of emitters and cavities. The labels in this figure correspond to the publications in the table in sec.~\ref{sec:appendix_3} of the appendix.}}
    \label{fig:expt_data}
\end{figure}

\onecolumngrid

In fig.~\ref{fig:expt_data} we track recent experimental progress in coupling quantum emitters to optical cavities (coupling strength $g$ and loss rates $\kappa$) for strong light-matter interactions. The labels in fig.~\ref{fig:expt_data} correspond to the publications listed in the table below.

\begin{center}
\begin{tabular}{ c c l }
 ~ & Year & Publication \\ 

S1 & 2007 & Controlling cavity reflectivity with a single quantum dot\\
S2 & 2008 & Controlled phase shifts with a single quantum dot\\
S3 & 2008 & Dipole induced transparency in waveguide coupled photonic crystal cavities\\
S4 & 2010 & Resonant excitation of a quantum dot strongly coupled to a photonic crystal nanocavity\\
S5 & 2010 & Fast electrical control of a quantum dot strongly coupled to a photonic-crystal cavity\\
S6 & 2012 & Ultrafast photon-photon interaction in a strongly coupled quantum dot cavity system\\
S7 & 2013 & A quantum logic gate between a solid-state quantum bit and a photon\\
S8 & 2013 & Strain tuning of a quantum dot strongly coupled to a photonic crystal cavity\\
S9 & 2013 & Coupling of a single nitrogen vacancy center in diamond to a fibre-based microcavity\\
S10 & 2015 & Polarization degenerate solid-state cavity quantum electrodynamics\\
S11 & 2016 & An integrated diamond nanophotonics platform for quantum-optical network\\
S12 & 2018 & Strongly cavity-enhanced spontaneous emission from silicon-vacancy centers in diamond\\
S13 & 2018 & Cavity-enhanced Raman emission from a single color center in a solid\\
S14 & 2018 & Photon-mediated interactions between quantum emitters in a diamond microcavity\\
S15 & 2019 & A gated quantum dot strongly coupled to an optical microcavity\\
S16 & 2020 & Experimental demonstration of memory-enhanced quantum communication\\
S17 & 2022 & Optical transparency induced by a largely purcell enhanced quantum dot in a polarization-degenerate cavity\\
S18 & 2023 & Entanglement of nanophotonic quantum memory nodes in a telecommunication network\\

\end{tabular}
\end{center}

\clearpage

\twocolumngrid
\subsection{Trajectories of Basis States during Training \label{sec:appendix_5}}

\begin{figure}[htbp]
    \centering
    \includegraphics[width = \columnwidth]{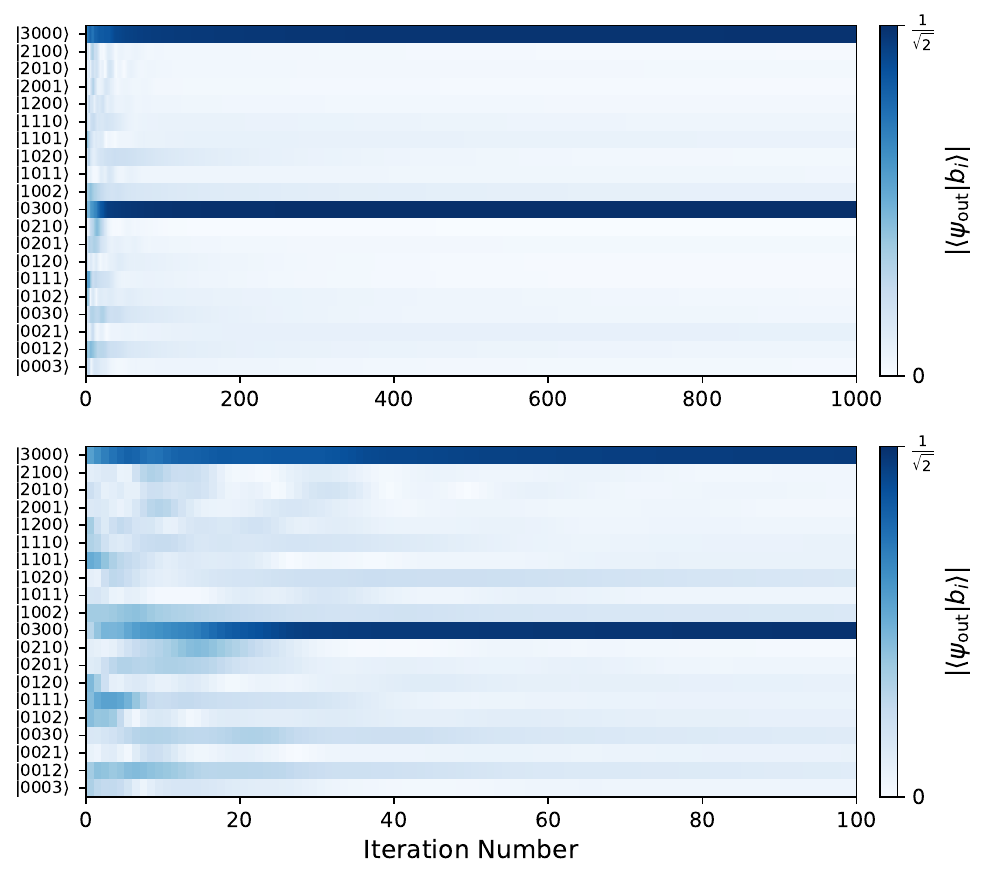}
    \caption{{\fontfamily{cmss}\selectfont\textbf{Evolution of the output state of the network while preparing a 3-photon $N00N$ state}. Amplitude of projection of output state of the network on each basis element as a function of iteration number. The upper plot indicates the fidelities over 1000 iterations, while the lower plot enlarges into the first 100 iterations. }}
    \label{fig:training_data}
\end{figure}

\begin{figure*}
    \centering
    \includegraphics[width = \columnwidth]{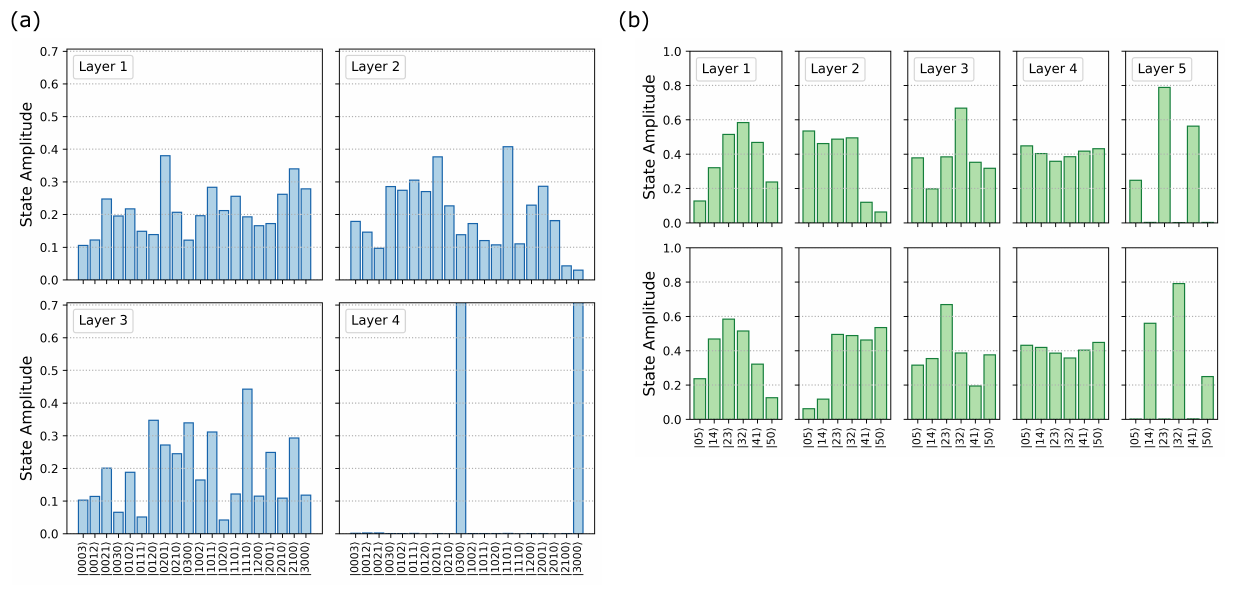}
    \caption{{\fontfamily{cmss}\selectfont\textbf{Trajectories of basis states during training}. \textbf{(a)} Amplitude of projection of output state of the network on each basis element, measured after every layer of the network to prepare the target 3-photon $N00N$ state. \textbf{(b)} Amplitude of projection of output state of the network on each basis element, measured after every layer of the network to prepapre the 5-photon two-mode $\chi^{(2)}$ binomial code. The upper row indicates the trajectories of the basis states when the input to the network is $\ket{0, 5}$, preparing the logical $\ket{\tilde{0}}$ state. The lower row indicates the trajectories of the basis states when the input to the network is $\ket{5, 0}$, preparing the logical $\ket{\tilde{1}}$ state. }}
    \label{fig:layer_data}
\end{figure*}

Sec.~\ref{sec:3} of the main text of the manuscript used the optimization routine to prepare multimode multi-photon states, and perform encoding and logical operations on bosonic codes. Here, we illustrate the trajectories of the basis state over the course of training. We demonstrate the results of preparation of a 3-photon $N00N$ state in a 4 mode network, i.e., $\ket{\psi_{\mathrm{target}}} = \frac{1}{\sqrt{2}} \left( \ket{3000} + \ket{0300} \right)$ in fig.~\ref{fig:training_data}(a). Each row plots the magnitude of the projection of the output state $\ket{\psi_{\mathrm{out}}}$ onto each basis element $\ket{b_{i}}$. The upper plot illustrates the fidelity over 1000 iterations of training. The lower plot enlarges into the first 100 iterations to illustrate the evolution of the populations of the other states. In fig.~\ref{fig:layer_data}, we plot the magnitude of the projection of the output state of each layer onto each basis element. The output state from the fourth layer is the target 3-photon $N00N$ state.

In fig.~\ref{fig:layer_data}(b), we plot the trajectories of the basis states in preparing the two-mode 5-photon $\chi^{(2)}$ binomial code. The upper row indicates the evolution of these basis states when the input to the network is $\ket{0, 5}$, to prepare the logical $\ket{\tilde{0}}$. The lower row indicates the evolution of these basis states when the input is $\ket{5, 0}$, to prepare the logical $\ket{\tilde{1}}$ state. 

\begin{figure*}
    \centering
    \includegraphics[width = \columnwidth]{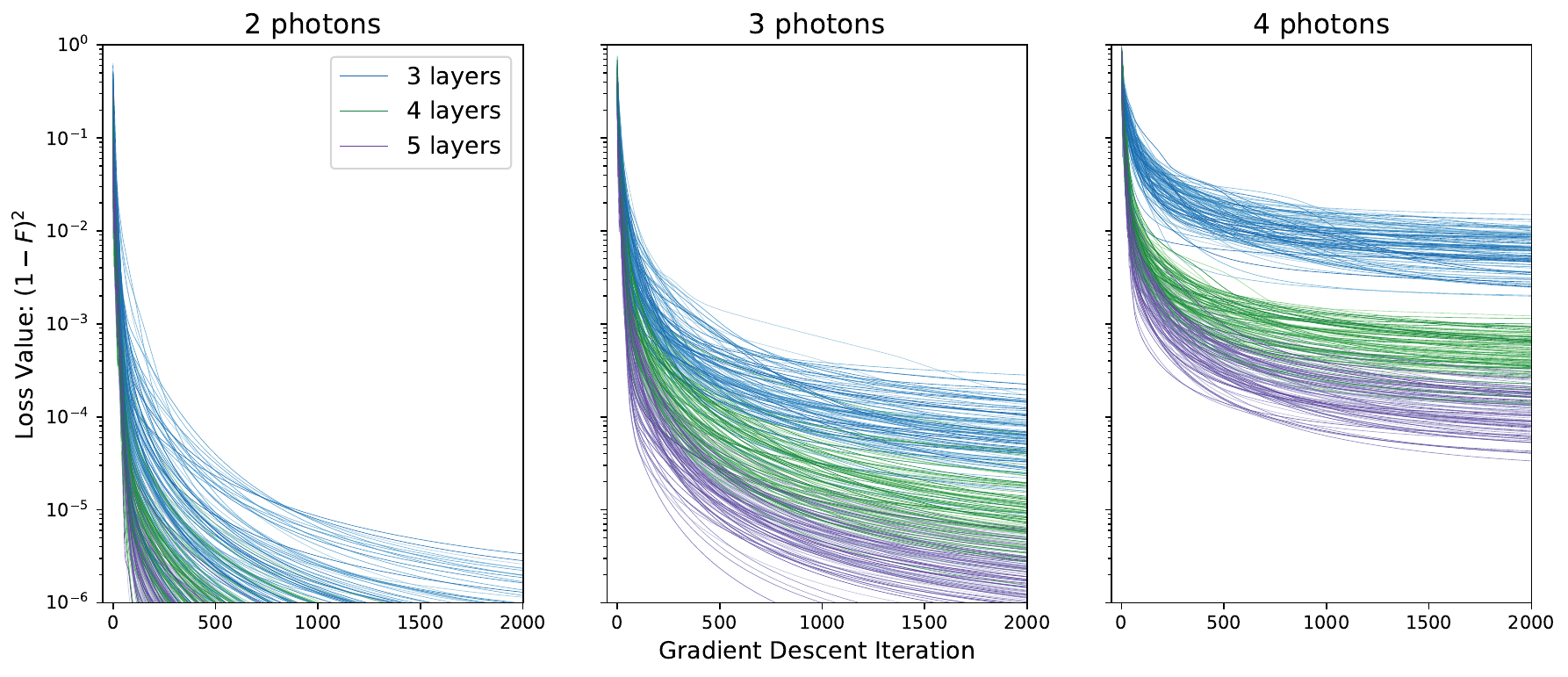}
    \caption{{\fontfamily{cmss}\selectfont\textbf{Loss as a function of iteration number in preparing Haar-random states}. Each facet depicts the loss after each step of the optimization routine in preparing the Haar-random states discussed in the main text of manuscript. Colors represent the depth of the network used to prepare a given target state. Increasing the depth of the network increases the achieved fidelity. As the number of photons increases (increasing the size of the Hilbert space), the optimization routine finds it increasingly difficult to reach high fidelities and begins to plateau. As we permit deeper networks, the loss further decreases and begins approaching the numerical noise floor.}}
    \label{fig:loss}
\end{figure*}

We also illustrate the loss as a function of iteration number is training the Haar-random states for 2-4 photons, using 3-5 layers in fig.~\ref{fig:loss}. Increasing the number of layers further decreases the loss.

\onecolumngrid

\twocolumngrid


\end{document}